\documentclass[tcfd]{svjour}
\usepackage{graphicx}
\usepackage{amsfonts}
              
\newcommand{\beq}{\begin{equation}}
\newcommand{\eeq}{\end{equation}}
\newcommand{\bea}{\begin{eqnarray}}
\newcommand{\eea}{\end{eqnarray}}
\newcommand{\tu}{\widetilde{u}}
\newcommand{\tS}{\widetilde{S}}
\newcommand{\ol}[1]{\overline{#1}}
\newcommand{\p}{\partial}
\newcommand{\R}{\mathbb{R}}
\newcommand{\Ord}{O}
\newcommand{\lap}{{\p_m^2}}
\newcommand{\fourth}{$4^\mathrm{th}$}
\newcommand{\second}{$2^\mathrm{nd}$}
\newcommand{\vct}[1]{\vec{#1}}
\newcommand{\cor}[2]{\rho({#1}^f_{#2},{#1}^t_{#2})}
\newcommand{\Err}[2]{\mathcal{E}({#1}^f_{#2},{#1}^t_{#2})}
\newcommand{\atanh}{\mathrm{atanh}}
\newsymbol\gtrsim 1326
\begin{document}

% watch the order of the \thanks: the svjour thing gets the order wrong
\title{Dynamic Model for LES Without Test Filtering: Quantifying the Accuracy 
of Taylor Series Approximations%
\thanks{%
This work has been supported financially by NSF-EAR 9909679 and NSF-CTS 9803385.}
}

%\renewcommand{\thefootnote}{\fnsymbol{footnote}}
%\footnotetext[1]{\rm This work has been supported financially by NSF-EAR 9909679 and NSF-CTS 9803385.}
%\renewcommand{\thefootnote}{\arabic{footnote}}

\Author{Stuart Chester, Fabrice Charlette%
\footnotemark[1]
%\thanks{%
%\emph{Permanent address:} Institut Fran\protect{\c{c}}ais du P\protect{\'e}trole, 1 et 4, avenue de Bois-Pr\protect{\'e}au, BP 311, 92506 Rueil-Malmaison, France.}
, and Charles Meneveau\mail{Telephone: +1 410 516 7802; Fax: +1 410 516 7254; e-mail: meneveau@jhu.edu}}{Department 
of Mechanical Engineering,\\ 
Center for Environmental and Applied Fluid 
Mechanics, \\
The Johns Hopkins University, 3400 N. Charles St., 
Baltimore, MD 21218, U.S.A.}

\footnotetext[1]{
\emph{Permanent address:} Institut Fran\protect{\c{c}}ais du P\protect{\'e}trole, 1 et 4, avenue de Bois-Pr\protect{\'e}au, BP 311, 92506 Rueil-Malmaison, France.}

\commun{Communicated by }
% Will be entered by Springer

\date{Received date and accepted date}

\abstract{%
The dynamic model for large-eddy simulation (LES) of turbulent flows requires 
test filtering the resolved velocity fields in order to determine model 
coefficients. However, test filtering is costly to perform in large-eddy 
simulation of complex geometry flows, especially on unstructured grids.
The objective of this work is to develop and test an approximate but less costly 
dynamic procedure which does not require test filtering. The proposed method is 
based on Taylor series expansions of the resolved velocity fields.  Accuracy is 
governed by the derivative schemes used in the calculation and the number of 
terms considered in the approximation to the test filtering operator.  The 
expansion is developed up to fourth order, and results are tested {\it a priori} 
based on direct numerical simulation data of forced isotropic turbulence in the context of the dynamic 
Smagorinsky model.  The tests compare the dynamic Smagorinsky coefficient 
obtained from filtering with those obtained from application of the
Taylor series expansion. They show that the expansion up to second order 
provides a reasonable approximation to the true dynamic coefficient (with 
errors on the order of about 5\% for $c_s^2$), but that including higher-order 
terms does not necessarily lead to improvements in the results due to inherent 
limitations in accurately evaluating high-order derivatives.  {\it A posteriori} 
tests using the Taylor series approximation in LES of forced isotropic 
turbulence and channel flow confirm that the Taylor series approximation yields 
accurate results for the dynamic coefficient.  Moreover, the simulations are stable 
and yield accurate resolved velocity statistics.}

\authorrunning{Chester, Charlette, and Meneveau}

\maketitle

\section{Introduction}
\label{intro}
In large-eddy simulation (LES), the Navier--Stokes equations  are filtered in 
an attempt to isolate the large-scale motion in a turbulent flow. The filtered 
equations contain the divergence of the sub-grid scale (SGS) stress, 
\beq
\label{eqn:def of SGS stress}
\tau_{ij}=\widetilde{u_i u_j} - \tu_i \tu_j,
\eeq
where ($\tilde{\ }$) denotes a filtering operation.  This filter, called the 
grid filter,  is a convolution with the kernel $G_\Delta$,
\beq
\label{eqn:def of grid filter}
\widetilde{f}(\vct{x}) = \int\limits_{\R^3} f(\vct{x}') G_{\Delta}(\vct{x},\vct{x}') \, \D\vct{x}',
\eeq
which has a smoothing action on the scales smaller than $\Delta$. Analogous to 
the closure problem related to the Reynolds-averaged Navier--Stokes equations,  
the SGS stress $\tau_{ij}$ must be modeled using available filtered quantities.  
A variety of SGS models exist \citep{Piomelli99,MeneveauKatz00}, but the 
results largely depend upon the choice of model coefficients, and the latter often 
have to be tuned from one flow regime to another. The dynamic procedure introduced 
by \citet{Germanoetal91} avoids such ad-hoc tuning. A crucial step in 
the dynamic procedure is using a filtering operation, called test filtering, to
gather information about the smallest resolved scales.   The test filtering 
operation, denoted by ($\bar{\ }$), is defined in a similar way to the grid 
filter, except the filter acts on a larger scale $\alpha \Delta$:
\beq
\label{eqn:def of test filter}
\ol{\widetilde{f}}(\vct{x}) = \int\limits_{\R^3} \widetilde{f}(\vct{x}') G_{\alpha \Delta}(\vct{x},\vct{x}') \, \D\vct{x}',
\qquad (\alpha > 1).
\eeq

In this paper, we restrict attention to the dynamic formulation 
\citep{Germanoetal91} of the Smagorinsky \citep{Smagorinsky63} eddy-viscosity 
model.  It approximates the deviatoric part of the SGS stress by
\beq
\label{eqn:Smagorinsky}
\tau_{ij} - \frac{1}{3} \tau_{kk} \delta_{ij} = - 2 (c_s \Delta)^2 |\tS| \tS_{ij},
\eeq
where $c_s$ is the Smagorinsky constant, $\tS_{ij} = \frac{1}{2}(\p_j \tu_i 
+ \p_i \tu_j)$  is the filtered rate of strain tensor, and 
$|\tS| = ( 2 \,\tS_{ij} \tS_{ij} )^{1/2}$ is the filtered strain-rate magnitude.
The expression for $c_s^2$, obtained by minimizing the mean square error in the 
Germano identity \citep[see][]{Germanoetal91,Lilly92}, 
$\mathcal{E} = \left\langle \left[L^{f}_{ij} - {\left(c_s^{f}\right)}^2 M^{f}_{ij}\right]^2 \right\rangle$, is
\beq
\label{eqn:c_s squared}
\left(c^{f}_s\right)^2 = \frac{ \langle L^{f}_{ij} M^{f}_{ij} \rangle }{ \langle M^{f}_{k \ell} M^{f}_{k \ell} \rangle },
\eeq
where the angle brackets denote an averaging operation \citep{Ghosaletal95},
\beq
\label{eqn:def of L}
L^{f}_{ij} = \ol{\tu_i \tu_j} - \ol{\tu}_i \ol{\tu}_j
\eeq
is the resolved turbulent stress, and
\beq
\label{eqn:def of M}
M^{f}_{ij} = 2 \Delta^2 \left[ \ol{|\tS| \tS_{ij}} - \alpha^2 |\ol{\tS}| \ol{\tS}_{ij} \right]
\eeq  
(the superscript ``$f$'' in both tensors refers to ``filtering''---to be later 
contrasted to Taylor series approximations). 
% begin Referee C
In (\ref{eqn:def of M}), for simplicity, we have put $\alpha$ in place of the 
ratio of the `compound' filter length to the grid filter length, where `compound' 
filter length is the effective length scale of the filter obtained by sequentially
applying the grid and test filters.  The error in doing so is tolerable for typical 
values of $\alpha$ and one can calculate the precise form of this ratio after 
choosing a specific type of test filter and assuming a form for the implicit grid 
filter. The reader is referred to \citet{Winckelmansetal98} for further 
discussion. The averaging in (\ref{eqn:c_s squared}) may
% end
be done over 
directions of statistical homogeneity \citep{Germanoetal91}, if any exist.  
In complex geometries without homogeneous directions, the Lagrangian dynamic
model \citep{Meneveauetal96}, which calculates time averages along pathlines, can 
be used.  In the present work, only turbulence with statistically homogeneous 
directions will be considered for simplicity. Hence, spatial averaging is 
employed in all applications.

However, from a practical perspective, test filtering adds computational cost 
to LES. The cost is typically manageable when dealing with pseudo-spectral 
numerical methods, where filter operations can be performed in Fourier space. 
When using physical-space based test filtering approaches (e.g. in 
finite-difference or finite-volume codes with structured grids), the operation 
count depends upon the number of neighboring grid-points involved in the 
filtering. 
% begin Referee C
\citet{NajjarTafti96} give a discussion of the effects of using test filters 
with finite-difference approximations and implications for the dynamic Smagorinsky
model.  
% end
When dealing with complex-geometry flows, unstructured grids are 
often employed for which one must decide which neighboring nodes are involved in 
the filtering. Challenges also arise when seeking parallelization. These 
difficulties have somewhat limited the applicability of the dynamic model to LES 
of complex-geometry flows. 
Various filtering operators for unstructured grids have already been proposed
and tested, in several papers by \citet{Jansen94,Jansen99}. He discusses 
and compares several options, including derivative based filtering,  
and generalized top-hat filtering \citep{Jansen99}. The generalized top-hat filter 
is a natural extension of the top-hat filter to unstructured grids by averaging 
over all elements that share a particular node.  
% begin Referee C
Filtering methods for complex geometries have also been discussed by 
\citet{MullenFischer99}.
% end

The derivative-based filter approximates the function to be test filtered by 
expanding it locally in a Taylor series and truncating.  This method is 
attractive because typical codes already have efficient methods in place to 
evaluate derivatives (as opposed to filtering, which is not typically needed in 
most existing codes). As reviewed in \citet{MeneveauKatz00}, the idea of 
expanding the local velocity field in a Taylor series has been used before in 
SGS modeling, mainly to simplify modeling terms for the similarity model and/or 
the Leonard stresses \citep{Leonard74,Clarketal79,Liuetal94,Winckelmansetal98}, 
and also in the context of so-called defiltering SGS models 
\citep{Stolz&Adams99, Kuertenetal99}.  
In the context of the dynamic Smagorinsky model, Taylor series expansion was 
originally suggested by \citet{GaoOBrien93} to analytically study the limit 
$\alpha \to 1$, but was not applied or extended in LES. 
As mentioned above, the 
derivative based filter (or Taylor series approach) was one of the options 
tested by \citet{Jansen99}, although no detailed results are presented for 
this approach. Jansen has used generalized top-hat filtering in his dynamic LES 
of airfoil flow  because this method has been found to be 
cheaper than the Taylor series method and yet gives similar 
accuracy \citep{Jansen99}, in the context of his specific code.  
% begin Referee C
The derivative based method has also been used to perform {\it a priori} studies of 
the sensitivity of various SGS models to the type of test filtering employed
\citep{SagautGrohens99}.
% end
%% begin Referee C
%See \citet{NajjarTafti96} for a dicussion of discrete test filters with finite 
%differences as applied to the dynamic Smagorinsky model, though they do not 
%discuss the derivative based method.
%% end

In the present paper we focus on the Taylor series method because it is a fairly 
general approach which can be formulated, studied, and tested in general, less 
code-specific terms.  Section \ref{sec:1} describes the formulation of the 
approach, in which the dynamic Smagorinsky coefficient is expressed entirely in 
terms of derivatives of the resolved velocity field and no test filtering is 
required. The accuracy of this approach is tested both {\it a priori} 
(\S~\ref{sec:2}) as well as {\it a posteriori} in LES of two benchmark flows, 
namely forced isotropic turbulence and minimal channel flow (\S~\ref{sec:3}). 
Structured grids are used both for the direct numerical simulation (DNS) used in 
the {\it a priori} tests and in the LES runs, in order to allow us to make 
comparisons in a highly controlled and standard numerical environment.
Basic conclusions are presented in 
\S~\ref{sec:4}. 

\section{Formulation}
\label{sec:1}

Given a quantity $\widetilde{f}(\vct{x})$ (time dependence is implicit) that is 
to be test filtered, $\widetilde{f}(\vct{x}')$ in Equation 
(\ref{eqn:def of test filter}) (written for a homogeneous filter) is replaced with 
its Taylor series expansion about the point $\vct{x}$.  This leads to simple 
integrations that are performed analytically. The result is 
(assuming uniform convergence)
\beq
\label{eqn:Taylor series}
\ol{\widetilde{f}}(\vct{x}) = \sum_{n=0}^{\infty} \frac{(-1)^n}{n!} \left. \p^{n}_{i_1 \cdots i_n} \widetilde{f} \right|_{\vct{x}} \langle y_{i_1} \cdots y_{i_n} \rangle,
\qquad \vct{y}=\vct{x}-\vct{x'},
\eeq
where now the angle brackets denote the operation of taking the mean with 
respect to a homogeneous filter $G_{\alpha \Delta}$, e.~g.,
\beq
\label{eqn:G-averaging}
\langle y_{i_1} \cdots y_{i_n} \rangle = \int\limits_{\R^3} y_{i_1} \cdots y_{i_n}
G_{\alpha\Delta}(\vct{y})\,\D\vct{y}.
\eeq
For an isotropic filter, we write $G_{\alpha \Delta}(\vct{y}) = 
G_{\alpha \Delta}(|\vct{y}|)$, and all terms with $n$ odd vanish, so that we can 
take $n=2m$.  The choice of an isotropic filter also effects the remaining terms 
since they all become isotropic tensors.  Since derivatives can only be 
calculated to a limited accuracy, and we are forced to truncate the Taylor 
series, this method yields an approximation to the filtering operation.  By 
varying the number of terms used in the Taylor series, and the way derivatives 
are calculated, the accuracy of this approximation can be varied. 

Here, a formulation based entirely on the Gaussian filter,
\beq
\label{eqn:Gauss}
 G^{\mathrm{Gauss}}_{\Delta}(\vct{y}) = \left( \frac{6}{\pi \Delta^2} \right)^{3/2}  \exp\left(-\frac{
6 |\vct{y}|^2 }{\Delta^2}\right),
\eeq
is given although the approach can be extended to any isotropic filter with 
finite moments  (this excludes the spectral cutoff filter, which has infinite 
second moments).  
% begin Referee A suggestion
Additionally, up to second order only, the following is also valid for the 
box filter, since the second moments of the Gaussian and box filters are the 
same. This means that the rest of the development in this section, up to 
second order only, can also be applied when using the box filter.  This fact 
is used later in \S 4.2, where the box filter is used to perform test 
filtering.
% end
Restricting attention to the Gaussian filter, the moments in 
(\ref{eqn:Taylor series}) can easily be evaluated as
\beq
\langle y_{i_1} \cdots y_{i_{2m}} \rangle = \sum \langle y_{i_{k_1}} y_{i_{k_2}} \rangle \cdots 
\langle y_{i_{k_{2m-1}}} y_{i_{k_{2m}}} \rangle,
\eeq
where the sum is over all $(2m)!/\left( 2^m m! \right)$ ways of partitioning 
$\left\{i_1, i_2,  \cdots , i_{2m} \right\}$ into pairs \citep{Isserlis18}.  
Using the result
\beq
\langle y_i y_j \rangle  = \frac{(\alpha \Delta)^2}{12} \delta_{ij},
\eeq
we have 
\beq
\label{eqn:moments}
\left< y_{i_1} \cdots y_{i_{2m}} \right> = \frac{(\alpha \Delta)^{2m}}{12^m} \sum \delta_{i_{k_1} 
i_{k_2}} \cdots \delta_{i_{k_{2m-1}} i_{k_{2m}}}.
\eeq
Using (\ref{eqn:moments}) in (\ref{eqn:Taylor series}), we have
\beq
\label{eqn:test filter operator}
\ol{\widetilde{f}}(\vct{x}) = \sum_{m=0}^\infty \frac{(\alpha \Delta)^{2m}}{24^m m!} \left(\nabla^2\right)^m 
\widetilde{f}(\vct{x}), 
\eeq   
where $\left(\nabla^2\right)^m$ denotes $m$ applications of the Laplacian 
operator.  This expression allows one to replace test filtered quantities that 
appear in the dynamic model by expressions involving resolved (grid-filtered) 
quantities and their derivatives.

Application of (\ref{eqn:test filter operator}) to $\tu_i$, $\tu_j$, and 
$\tu_i \tu_j$  yields the approximation
\beq
\label{eqn:approx to L}
L^{t}_{ij}  =  \frac{(\alpha \Delta)^2}{12} {\frac{\p \tu_i}{\p x_k} \frac{\p \tu_j}{\p x_k}}  
  + \frac{(\alpha \Delta)^4}{288} \left[ \frac{\p \tu_i}{\p x_k} \frac{\p}{\p x_k}\left( \frac{\p^2 \tu_j}{\p x_m^2} \right) + \frac{\p^2 \tu_i}{\p x_k \p x_\ell} \frac{\p^2 \tu_j}{\p x_k \p x_\ell} + \frac{\p}{\p x_k} \left( \frac{\p^2 \tu_i}{\p x_m^2} \right) \frac{\p \tu_j}{\p x_k} \right] + \Ord(\Delta^6).
\eeq
Here, the terms involving derivatives of the product $\tu_i \tu_j$ have been 
expanded to take advantage of the favorable cancellation.  The superscript ``$t$'' 
stands for ``Taylor series approximation''. Similarly, after 
applying (\ref{eqn:test filter operator}) to $\tS_{ij}$ and $|\tS|\tS_{ij}$,
we can calculate an approximation for $M_{ij}$,  
\bea 
\label{eqn:approx to M}
M^{t}_{ij} &=& 2 \Delta^2 \left\{ \left( |\tS|\tS_{ij}-\alpha^2|{\tS^t}|\tS_{ij}\right)+ 
\frac{(\alpha\Delta)^2}{24} \left[ \lap\left(|\tS|\tS_{ij}\right) -
\alpha^2|{\tS}^t|\lap\tS_{ij}\right] \right. \nonumber \\
       & & \left. + \frac{(\alpha\Delta)^4}{1152}\left[ \nabla^4\left(|\tS|\tS_{ij}\right) -  \alpha^2
|{\tS}^t| \nabla^4\tS_{ij} \right] + \Ord(\Delta^6)\right\},
\eea
where $|{\tS}^t|$ is the derivative-based approximation to $|\ol{\tS}|$,
\beq
\label{eqn:mag S}
|{\tS}^t| = \left[ 2 \left(\tS_{ij} + \frac{\left(\alpha \Delta \right)^2}{24} \lap \tS_{ij} + \frac{\left(\alpha \Delta \right)^4}{1152}\nabla^4 \tS_{ij} + \Ord(\Delta^6) \right)^2 \right]^{1/2}.
\eeq
Here, the derivatives of products have not been expanded, since no cancellation 
in the terms of $M^{t}_{ij}$ occurs. The dynamic coefficient $c^{t}_s$ can then 
be computed as before by evaluating tensor contractions and averaging over 
homogeneous directions.  
For example, below we will make frequent use of the case where only the \second\ 
order expansions are kept, so that we have
\beq 
\label{eqn:c_s Taylor}
{c^t_s}^2 = \frac{\alpha^2}{24} \frac{ \left< ({\p_k \tu_i \p_k \tu_j}) 
\left( \left( |\tS|-\alpha^2|{\tS}^t| \right) \tS_{ij}  
 + \frac{(\alpha\Delta)^2}{24} \left[ \lap \left(|\tS|\tS_{ij}\right) -
\alpha^2 |{\tS}^t| \lap \tS_{ij} \right] \right) \right>}
{ \left< \left( \left(|\tS|-\alpha^2|{\tS}^t| \right) \tS_{\ell n}+
\frac{(\alpha \Delta)^2}{24} \left[ \lap \left(|\tS|\tS_{\ell n}\right) -
\alpha^2|{\tS}^t|\lap\tS_{\ell n}\right] \right)^2\right>},
\eeq
with
\beq
\label{eqn:mag S explicit} 
|{\tS}^t| = \left[ 2 \left(\tS_{ij} + \frac{\left(\alpha \Delta \right)^2}{24} \lap \tS_{ij} \right)^2 \right]^{1/2}.
\eeq

In the case of isotropic turbulence, the test filtering is done in three 
dimensions, so that the values of the indices $k$ and $m$ in 
(\ref{eqn:c_s Taylor}) are $1$, $2$, $3$, and the averaging in the numerator and 
denominator is volume averaging.  However, in the case of channel flow, in order 
to approximate planar (in $x$-$z$ planes) instead of volumetric filtering, 
the Laplacian term in the expansion of (\ref{eqn:Taylor series}) only contains 
derivatives in the $x$ and $z$ directions (i.~e. the indices $k$ and $m$ in 
(\ref{eqn:c_s Taylor}) and (\ref{eqn:mag S explicit}) only cycle over the 
two values $k=1$, $3$ and $m=1$,$3$).  Additionally, in the channel flow case, 
plane averaging  of the numerator and denominator of (\ref{eqn:c_s Taylor}) is 
performed in the $x$-$z$ planes. 

%%%%%%%%%%%%%%%%%%%%%%%%%%%%%%%%%%%%%%%%%%%%%%%%%%%%%%%%%%%%%%%%%%%%%%%%%%%%%
\section{{\em A priori} tests}
\label{sec:2}

\subsection{Methods}
\label{sec:2.1}

The {\it a priori} tests are done using a $128^3$ DNS database of forced 
isotropic turbulence at $R_{\lambda} \approx 94$, produced 
using the same pseudo-spectral algorithm as in \citet{Ceruttietal98}.  The grid 
filter size, $\Delta$, is chosen to be the grid spacing of a $32^3$ grid, and 
corresponds to $\Delta/\eta \sim 9$, where $\eta$ is the Kolmogorov scale. 
To perform the data analysis on the $32^3$ 
grid,  the $128^3$ DNS data is  grid filtered (using the Gaussian filter) and 
stored back on the DNS grid.  Then, the grid-filtered velocity is filtered using 
a spectral cutoff filter of width $\Delta$ (the same as the $32^3$ grid spacing) 
so that no aliasing occurs when transferring the velocity to the $32^3$ grid.  
The grid-filtered velocity, on the coarse grid, is then used to calculate the 
filtered rate of strain tensor using finite differences.  The model coefficient 
$c_s^2$ is calculated using the approximations (\ref{eqn:approx to L}), 
(\ref{eqn:approx to M}), and (\ref{eqn:mag S}), and is compared to values 
obtained by explicitly applying  a test filter.  
The series (\ref{eqn:test filter operator}) must be truncated, 
so here two levels of accuracy are considered: second order and fourth order.  The 
second order approximation (denoted as $T=2$) is obtained by keeping terms
$\Ord(\Delta^2)$ and lower in (\ref{eqn:test filter operator}), so only the first 
term in (\ref{eqn:approx to L}), and only the first two terms inside the 
curly braces in (\ref{eqn:approx to M}) are kept. The fourth order approximation 
(denoted as $T=4$) is obtained in an analogous way and contains all of 
the explicitly shown terms in (\ref{eqn:approx to L}) and (\ref{eqn:approx to M}).  

At both levels of accuracy, all of the derivatives required  to implement the 
derivative-based test filtering are initially calculated using centered finite 
differences accurate to second order (denoted as $D=2$).  These derivatives 
could be calculated spectrally, however the goal here is to examine how the 
derivative-based method performs when derivatives are calculated as they would 
be in a complex geometry situation. In an attempt to separate the errors due to 
truncating the series (\ref{eqn:test filter operator}) and finite differencing 
errors, fourth order finite differences (denoted as $D=4$) are also tried for 
calculating derivatives. This 
allows one to keep the filter size constant while the accuracy of the finite 
differences is changed.  

A coarse $32^3$ grid is used to store the grid-filtered velocity during the 
{\it a priori}  test, so that all subsequent calculations would be 
the same as if a LES was actually being performed.  This is also consistent with 
the choice of grid filter size.  The ratio of the widths associated with the 
test filter and the grid filter, $\alpha$, is varied between $2$, $3$, and $4$,
though in this study, the most attention is given to the common choice $\alpha=2$. 

Finally, the approximations (\ref{eqn:approx to L}) and
(\ref{eqn:approx to M}) are calculated to the desired level of accuracy and the 
results are used in (\ref{eqn:c_s squared}) to calculate $c_s^2$.  Spatial 
averaging is employed in (\ref{eqn:c_s squared}) because of the spatial 
homogeneity of the velocity field under consideration.  Note that since 
(\ref{eqn:approx to L}) and (\ref{eqn:approx to M}) contain products of velocity 
components, they  also contain aliasing errors.  However, these were not removed 
since it is assumed that methods to do so may not be available during a typical LES 
of a complex geometry flow.  
  
\subsection{Results}
\label{sec:2.2}
The results obtained from using the Taylor series approximations are compared to 
those obtained by performing the classical test filtering. The comparison is 
based on scatter plots of tensor element values for some representative cases, 
and on correlation coefficients and normalized mean square errors among 
individual tensor elements.  The correlation coefficient for a given tensor 
element is defined as
\beq
\rho(L^{t}_{mn},L^{f}_{mn}) = 
\frac{\left< L^{t}_{mn} L^{f}_{mn} \right> - \left< L^{t}_{mn} \right> \left< L^{f}_{mn} \right>}
{\left[\left(\left< {L^{t}_{mn}}^2 \right> - \left< L^{t}_{mn} \right>^2\right)~\left(\left< {L^{f}_{mn}}^2 \right> - \left<{L^{f}_{mn}} \right>^2\right)\right]^{1/2}}
\eeq 
(no summation over indices), and similarly for the $M_{ij}$ tensor elements. 
In order to also quantify the agreement among the magnitudes of the tensors, we 
compute the normalized square error \citep{Liuetal99} defined as
\beq 
\mathcal{E}(L^{t}_{mn},L^{f}_{mn})=\frac{\langle (L^{t}_{mn}  - L^{f}_{mn})^2 \rangle}{\langle
(L^{f}_{mn})^2 \rangle},
\eeq 
and similarly for the $M_{ij}$. Also the dynamic coefficient $c_s^2$ is 
evaluated by averaging the tensor contractions, and the relative error in 
$c_s^2$ is obtained.  

\subsubsection{Second order approximation}
\label{sec:2.2.1}
\begin{figure*}[p]
\centerline{%
\begin{tabular}{c@{\hspace{6pc}}c}
\includegraphics[width=2.5in]{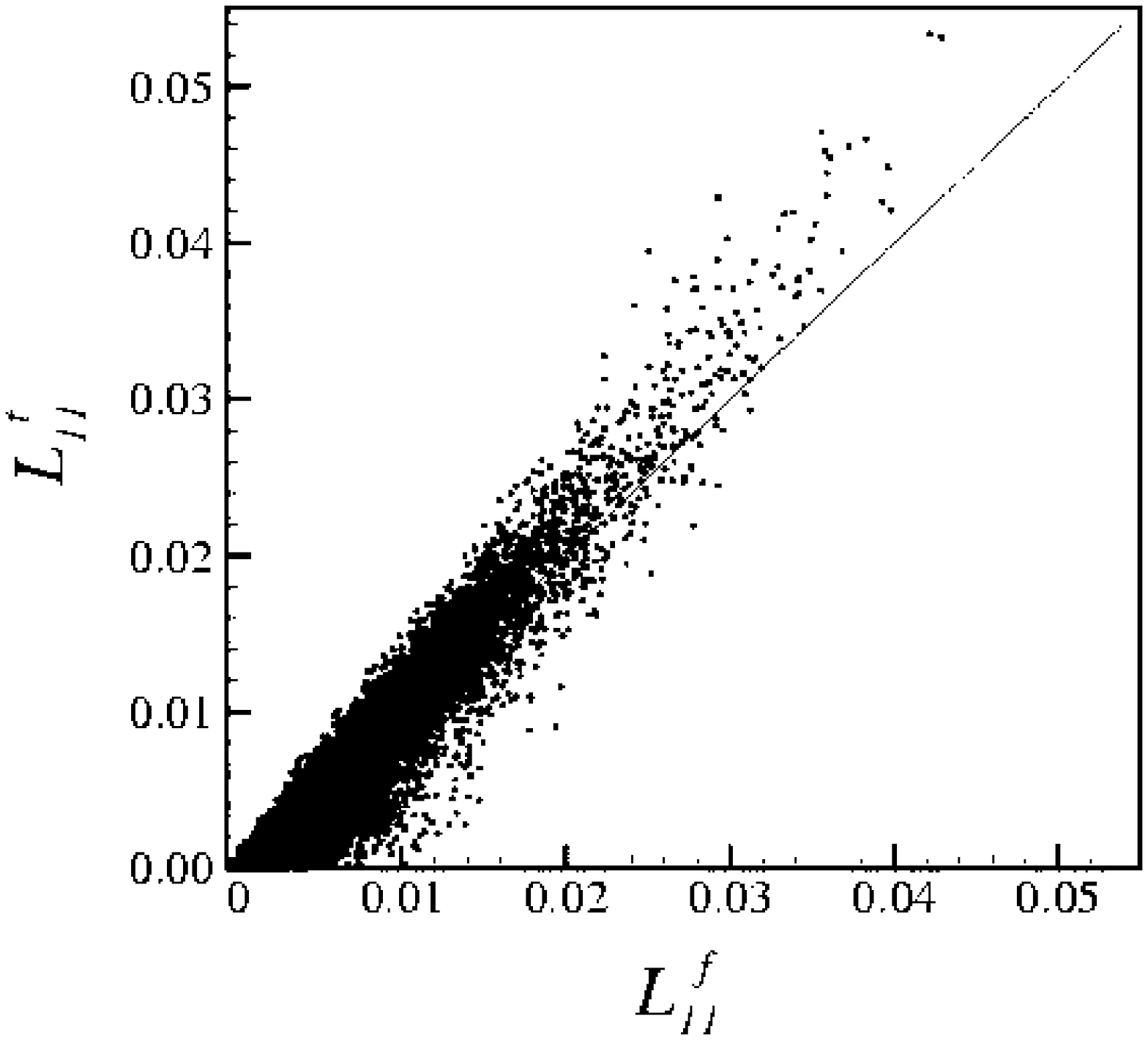} &
\includegraphics[width=2.5in]{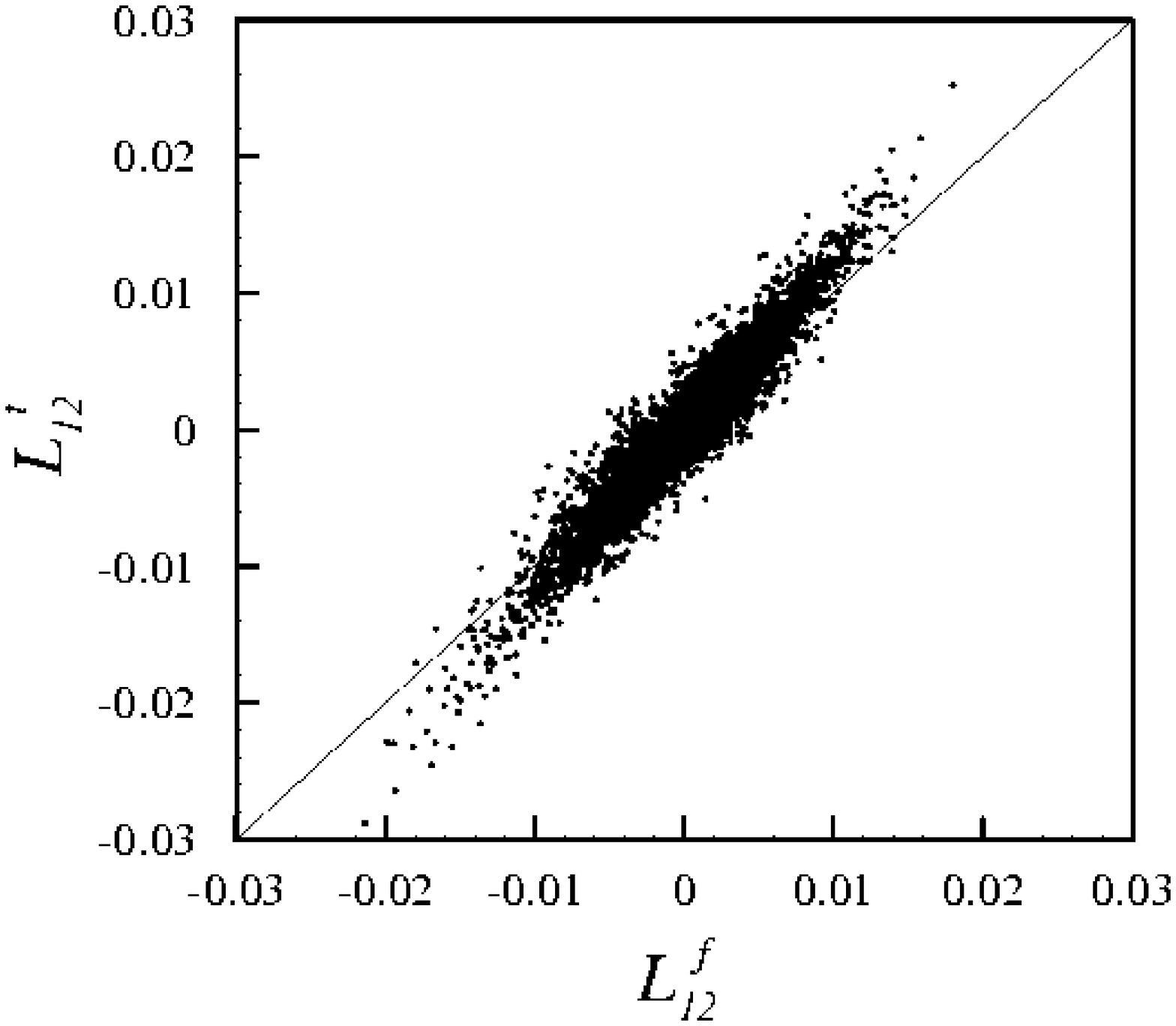} \\
(a) & (b)
\end{tabular}}
%\centercap
\caption{Scatter plot of typical components of $L_{ij}$ evaluated using the Taylor series approach compared to the classical filter-based approach, for the case $\alpha=2$ with the \second\ order approximation ($T=2$). (a) shows $L_{11}$ while (b) shows $L_{12}$.} 
\label{fig:L2scatter}
\end{figure*}

In this section, the results obtained by using the second order approximation 
($T=2$) and second order finite differencing ($D=2$) are presented. Here, the parameter $\alpha$ is fixed at the 
common value of 2; effects of varying $\alpha$ are discussed in 
\S~\ref{sec:1.2.5}. First, we compare the results for the components 
$L_{ij}$ obtained using the two methods.   Scatter plots comparing the \second\ 
order approximation to the exact (filter-based) results are shown in 
Figure~\ref{fig:L2scatter} for a diagonal and an off-diagonal component of $L_{ij}$.
Both plots indicate that the \second\ order approximation is overall quite good, 
but has a tendency to under-estimate the magnitude of the $L$-components for small 
$|L_{ij}|$, but over-estimates the magnitude at larger $|L_{ij}|$. 

As shown in the first column of Table~\ref{table:results}, the \second\ order 
approximation is well correlated ($\rho \sim 0.95$) with the exact results, and 
the normalized square error between the two is $\sim 6 \%$ for diagonal terms 
and $\sim 15 \%$ for off-diagonal terms.

\begin{figure*}[p]
\centerline{%
\begin{tabular}{c@{\hspace{6pc}}c}
\includegraphics[width=2.5in]{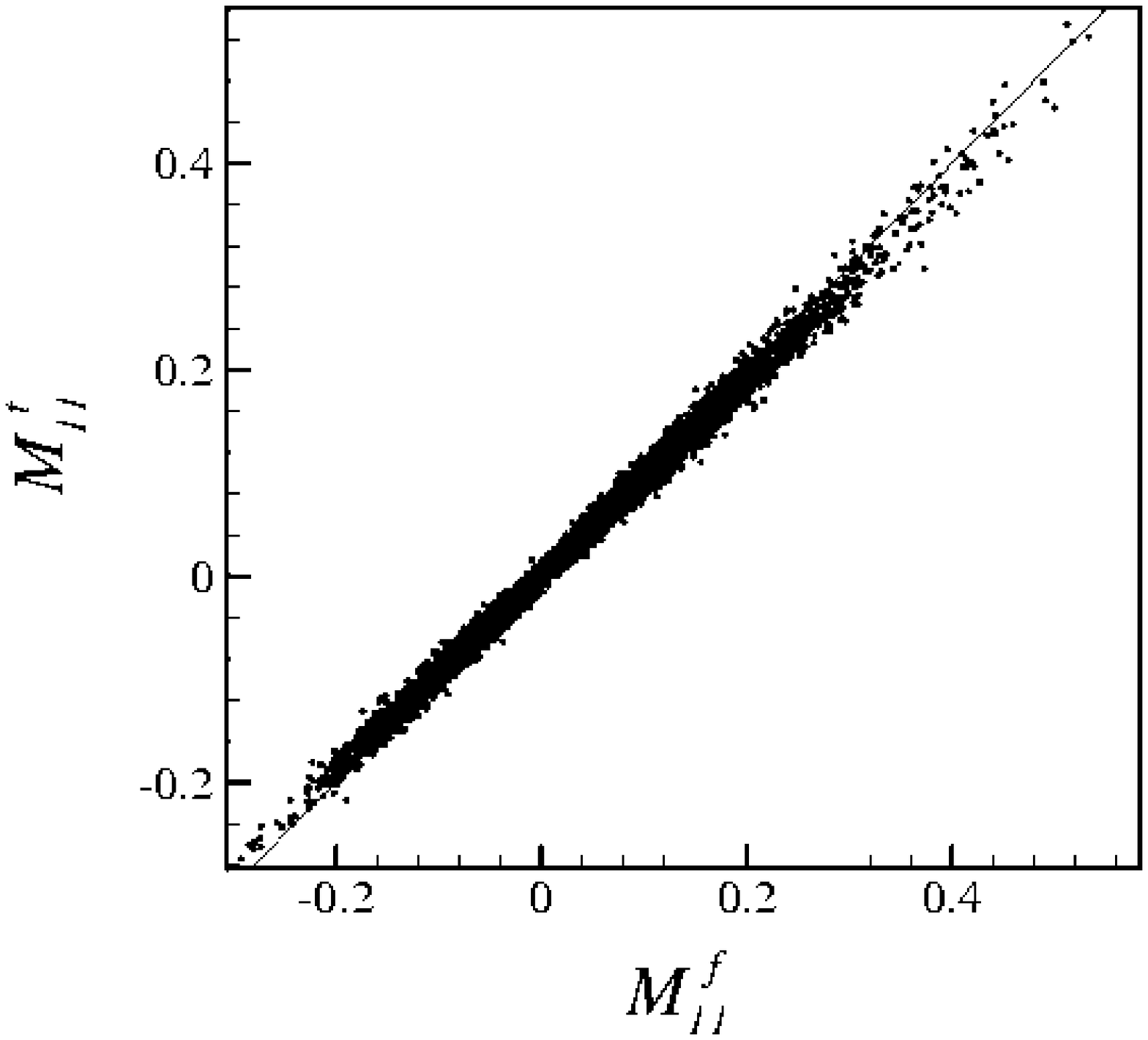} &
\includegraphics[width=2.5in]{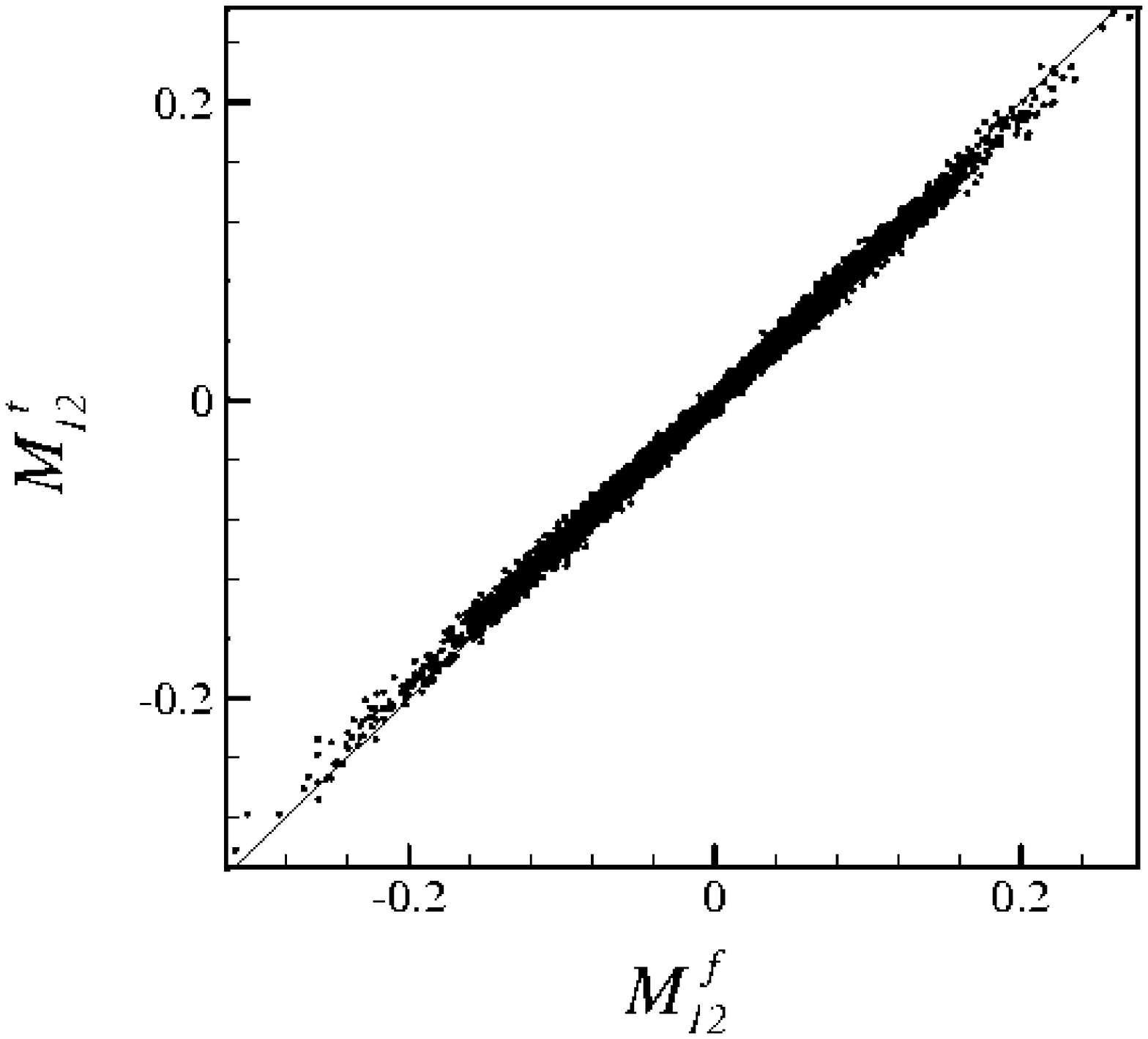} \\
(a) & (b)
\end{tabular}}
%\centercap
\caption{Same as Figure~\ref{fig:L2scatter}, but for the tensor $M_{ij}$.(a) shows $M_{11}$ while (b) shows $M_{12}$.}
\label{fig:M2scatter}
\end{figure*}

Scatter plots of components of $M_{ij}$ obtained using the \second\ order 
approximation and the exact results are shown in Figure~\ref{fig:M2scatter}, from 
which it can be seen that the approximation is very good. As shown in 
Table~\ref{table:results}, the \second\ order approximation for the $M_{ij}$ is very 
highly correlated with the exact values, with $\rho \sim 0.99$.  The normalized 
mean square is significantly lower than it was for the $L_{ij}$, being 
$\sim 1 \%$ for diagonal terms and $\sim 0.5 \%$ for off-diagonal terms.   

Using the  approximate and exact results for $L_{ij}$ and $M_{ij}$ in 
Equation (\ref{eqn:c_s squared}), and using volume averaging gives the values of 
$c_s^2$ shown in the last entries of Table~\ref{table:results}.  The relative 
error in $c_s^2$ obtained by using the \second\ order approximation is $\sim 5 \%$.  
% begin Referee C
Since the value of the coefficient obtained in the present work is essentially 
the same as for the standard filtering approach, the dissipation characteristics 
will be the same as for the standard filter-based dynamic model.
% end

\subsubsection{Fourth order approximation}
\label{sec:2.2.2}
\begin{figure*}[p]
\centerline{%
\begin{tabular}{c@{\hspace{6pc}}c}
\includegraphics[width=2.5in]{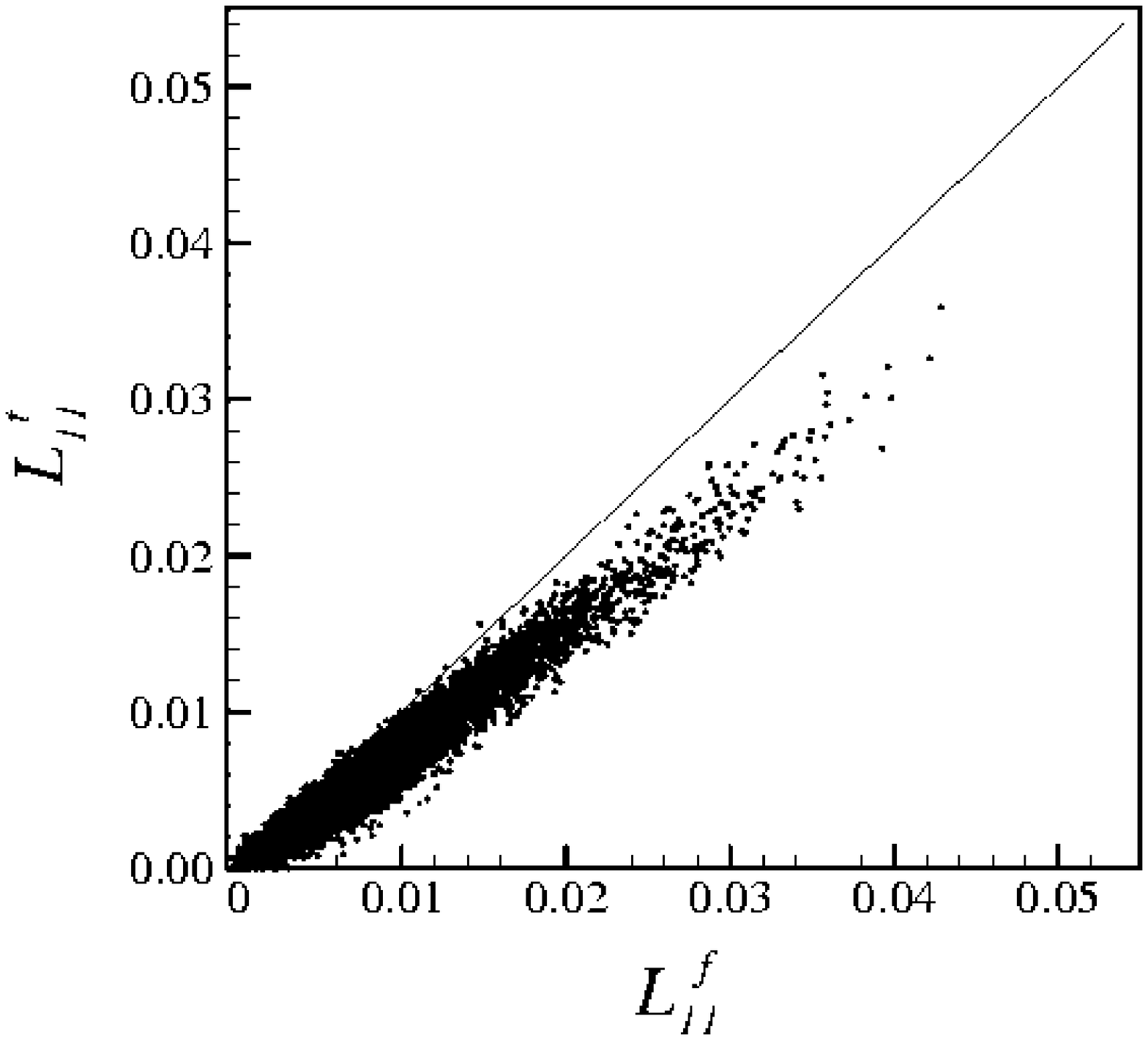} &
\includegraphics[width=2.5in]{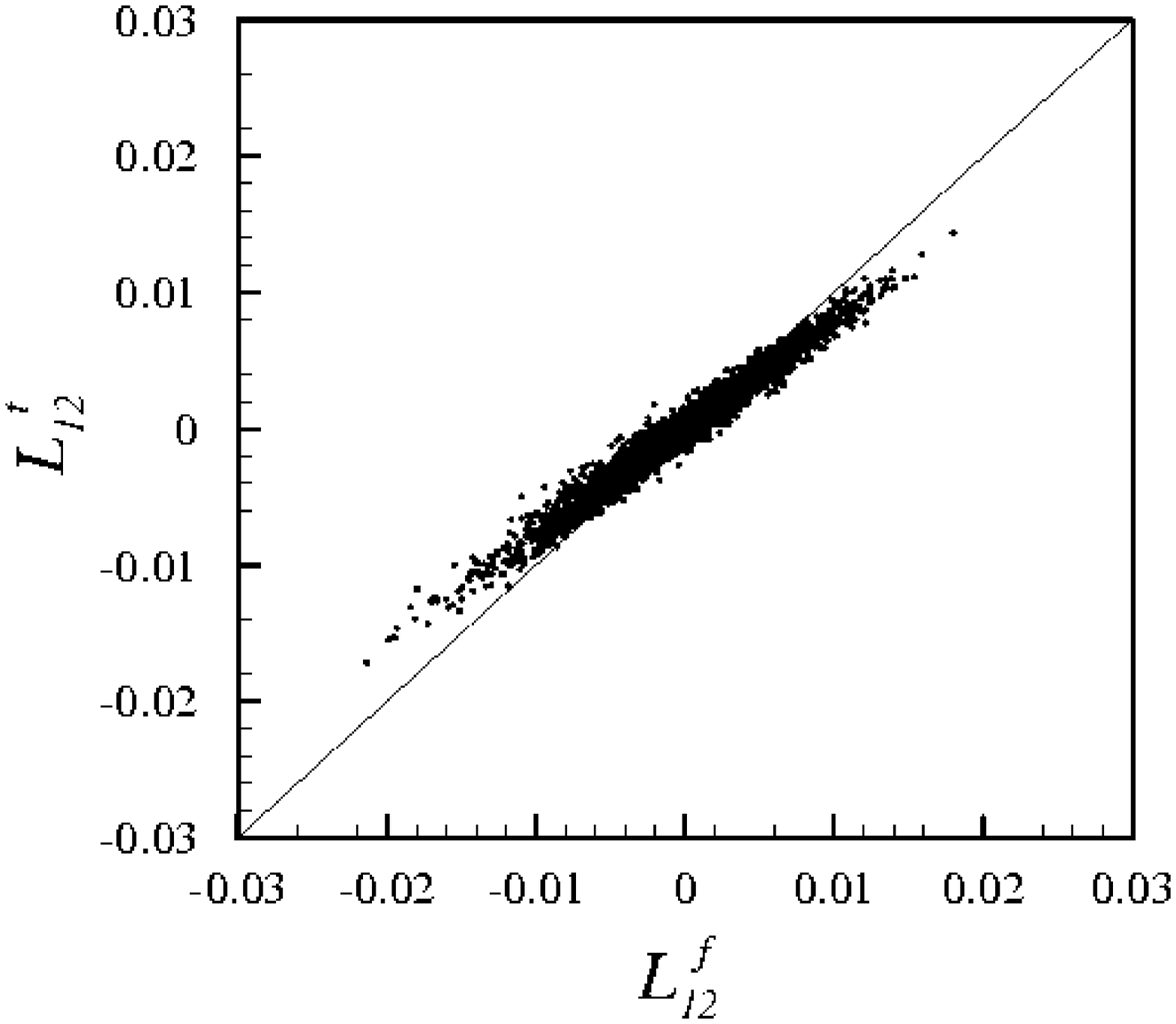} \\
(a) & (b)
\end{tabular}}
%\centercap
\caption{Same as Figure~\ref{fig:L2scatter} for the case $\alpha=2$, but now with the \fourth\ order approximation ($T=4$). (a) shows $L_{11}$ while (b) shows $L_{12}$.} 
\label{fig:L4scatter}
\end{figure*}

To answer the question of whether one can improve on the $5 \%$ relative error 
of the \second\ order approximation by including higher order terms, {\it a 
priori} tests are done with the \fourth\ order approximation ($T=4$, 
but still $D=2$).  As in the \second\ order case above, we fix $\alpha=2$ here.
Beginning with the components of $L_{ij}$, the scatter plots in 
Figure~\ref{fig:L4scatter} show that the \fourth\ order approximation tends to 
under-estimate the magnitude of the components of $L_{ij}$.   Despite this, the 
correlation coefficient between the \fourth\ order and exact results, shown in 
the second column of Table~\ref{table:results} to be
$\sim 0.98$, is higher than it was between the \second\ order and exact results.
Additionally, the normalized square error between the approximation and exact 
results for the $L_{ij}$, also in Table~\ref{table:results}, is lower overall 
with an improvement in the off-diagonal terms outweighing a slight increase in 
the error of the diagonal terms.  Note that with the use of the \fourth\ order 
approximation, there is no longer a large discrepancy between the diagonal and 
off-diagonal terms, as was observed when the \second\ order approximation was 
used, as the normalized square error is $\sim 7 \%$.

\begin{figure*}[p]
\centerline{%
\begin{tabular}{c@{\hspace{6pc}}c}
\includegraphics[width=2.5in]{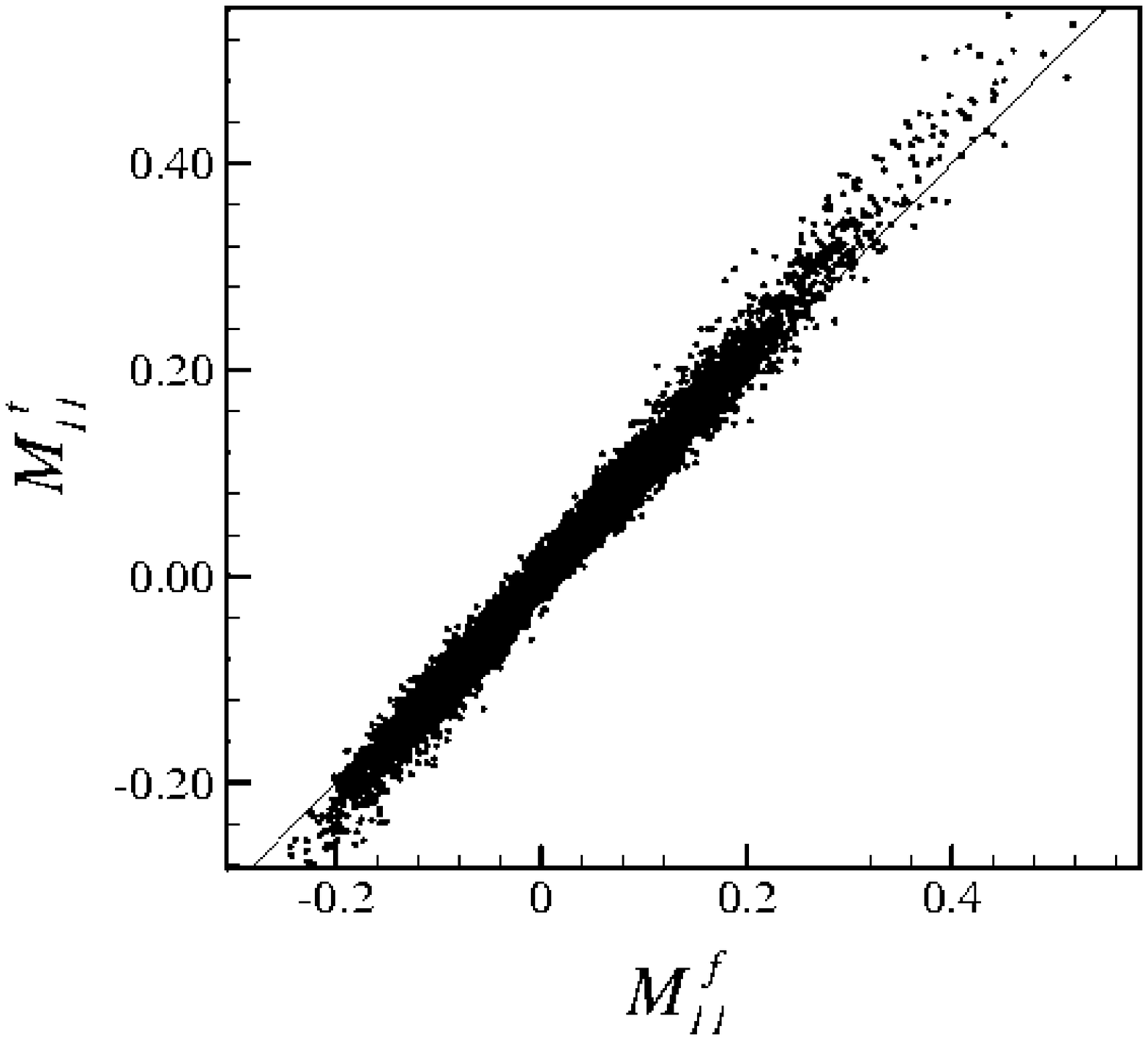} &
\includegraphics[width=2.5in]{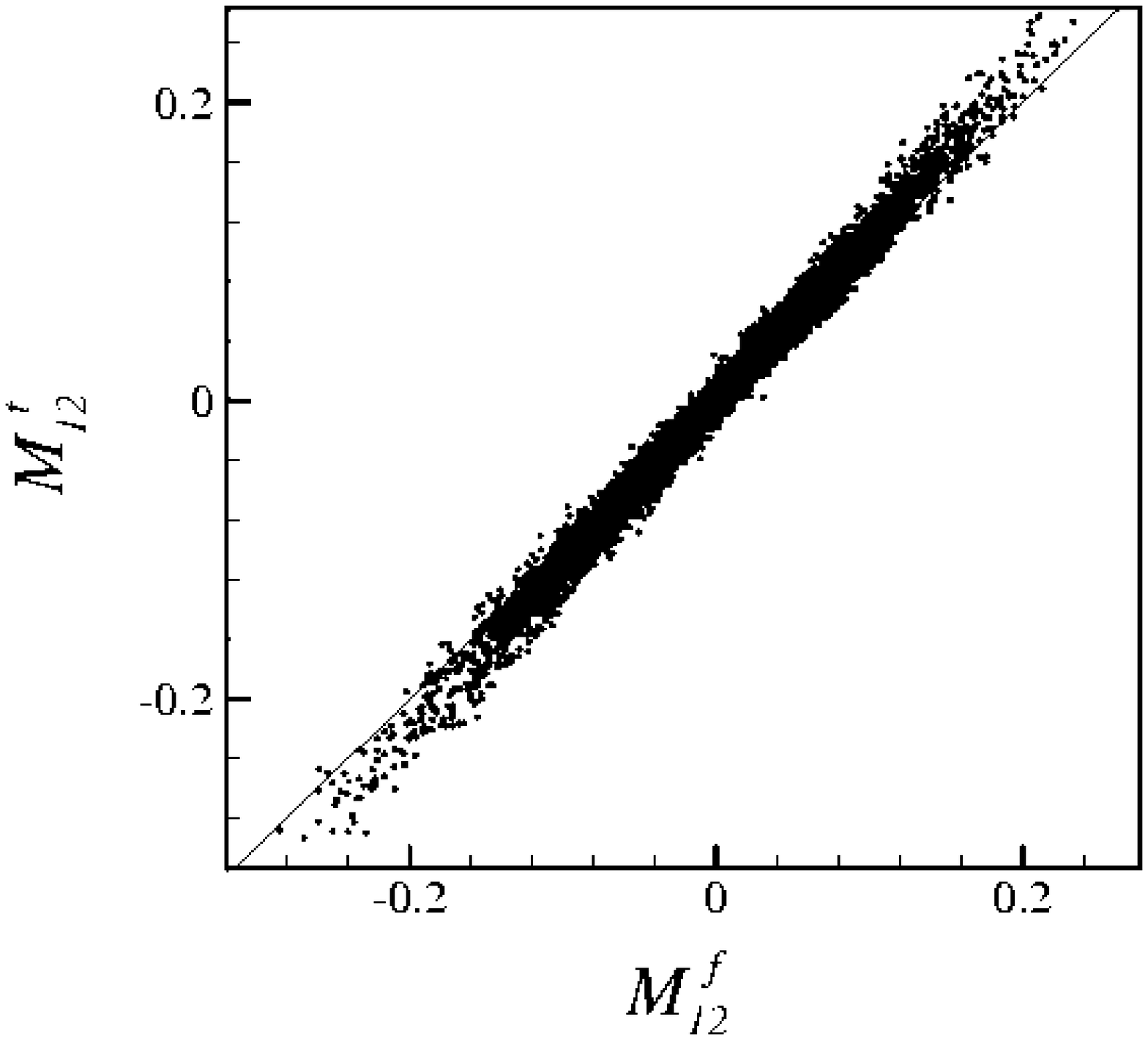} \\
(a) & (b)
\end{tabular}}
%\centercap
\caption{Same as in Figure~\ref{fig:M2scatter} using the the \fourth\ order approximation ($T=4$). (a) shows $M_{11}$ while (b) shows $M_{12}$.} 
\label{fig:M4scatter}
\end{figure*}

For $M_{ij}$, Figure~\ref{fig:M4scatter} shows  scatter plots obtained with the 
\fourth\ order approximation, which now shows a slight tendency to over-estimate 
the magnitude of $M_{ij}$.  The correlation coefficient is seen in table
\ref{table:results} to remain high, at $\sim 0.99$.  However, the normalized 
square error for the $M_{ij}$ components is increased significantly to 
$\sim 3 \%$ for both diagonal and off-diagonal terms by using the \fourth\ order 
approximation.  This is an increase from $1 \%$ for diagonal terms and 
$0.5 \%$ from the off-diagonal terms, which were obtained with the \second\ 
order approximation.  As a result, the value of $c_s^2$ obtained with the 
\fourth\ order approximation has a relative error of $\sim 27 \%$.  This is 
significantly worse than the results obtained using the \second\ order 
approximation.     

There are two likely reasons why the \fourth\ order approximation gives 
inferior  results. The first is that the finite difference errors in the 
\second\ order terms are swamping the \fourth\ order correction. In other words, 
due to the second-order finite differencing ($D=2$), the error in 
$\p_k \tu_i \p_k \tu_j$ in Equation~\ref{eqn:approx to L} is $\Ord(\Delta^4)$---the same 
order as the \fourth\ order correction.  The second possible reason is that the 
filtered velocity field is simply not smooth enough to allow reliable 
calculation of high order derivatives through finite differencing. This issue is 
an inherent, well-known difficulty associated with LES velocity fields that have 
a Kolmogorov energy spectrum. Even after removing some energy near the 
grid-scale through the implicit grid filter, gradients are still dominated by modes 
very near the scale $\Delta$.  Also note that the main problems seem to be 
associated with the error in the \fourth\ order $M_{ij}$ terms, which contain 
the highest order velocity derivatives.  These two possibilities are examined in 
more detail below.   

\subsubsection{Higher order derivative scheme}
\label{sec:2.2.3}
To check the possibility of whether the truncation error from the 
$\Ord(\Delta^2)$ finite differences is swamping the \fourth\ order correction, 
more {\it a priori} tests were done using $\Ord(\Delta^4)$ finite differences 
to evaluate the \second\ order terms.  This ensures that the finite difference
truncation error from the \second\ order terms is now of higher order than 
the \fourth\ order correction terms.
% begin Referee C
To compare the relative importance of finite differencing error and the 
error associated with truncation in the Taylor series method, we consider  
the form of the leading error terms. This is done in one dimension only for 
simplicity.  In the case $T=D=2$, the expansion with finite differences is
\beq
\label{eqn:T2D2}
\overline{\widetilde{f}}(x) = 
\widetilde{f}(x) + \frac{(\alpha\Delta)^2}{24}
\left( \widetilde{f}''(x) \right)_{fd} + {\alpha^2\Delta^4} \left( 
\frac{\alpha^2-2}{1152} \right) \left( \widetilde{f}^{iv}(x) \right)_{fd} 
+ O(\Delta^6),
\eeq
where the subscript $fd$ refers to derivatives calculated with centered finite 
differences, e.g. $\left( \widetilde{f}''(x) \right)_{fd} 
= (\widetilde{f}(x+\Delta) - 2 \widetilde{f}(x) + \widetilde{f}(x-\Delta))/
\Delta^2$. In the case $T=4$, $D=2$, 
the leading error term is the $2/1152$ part of the coefficient of 
$\widetilde{f}^{iv}$ in the above expression, which is of 
the same order (in $\Delta$) as the fourth order term in the approximation to the 
test filter (i.e. the $\alpha^2/1152$ part of the coefficient of 
$\widetilde{f}^{iv}$). In particular, when $\alpha = 2$ the 
finite differencing error is smaller but comparable to the fourth order correction 
term in the Taylor series expansion for the test filtered quantities. 

With $D=4$, the expansion becomes
\beq
\label{eqn:T4D4}
\overline{\widetilde{f}}(x) = \widetilde{f}(x) + \frac{(\alpha\Delta)^2}{24} 
\left( \widetilde{f}''(x) 
\right)_{fd} + \frac{(\alpha\Delta)^4}{1152} \left( \widetilde{f}^{iv}(x) 
\right)_{fd}  
+ \alpha^2 \Delta^6 \left(\frac{ 5 \alpha^4 + 192}{414720} \right)
\left( \widetilde{f}^{vi}(x) \right)_{fd} + O(\Delta^8).
\eeq
In the case $T=2$, $D=4$, we see that the leading error term is simply the 
fourth order correction term in the Taylor series expansion of the test 
filtering operator (the $(\alpha\Delta)^4/1152$ term).  While in the case 
$T=D=4$, the leading error term is of higher order than the fourth order 
correction in the expansion of the test filtering operator (it is $\propto 
\Delta^6$). When $\alpha=2$, the finite difference error is larger, though of 
the same order of magnitude as the correction term for the test filter expansion.

% end   
The results for the $L_{ij}$ and $M_{ij}$ 
components, again for $\alpha=2$, are summarized in the third  and fourth 
columns of Table~\ref{table:results}.   For the \second\ order approximation, 
although the correlation coefficients remain nearly the same as when the 
\second\ order finite differences were used, the normalized square error for 
$L_{ij}$ shows a significant increase, while it shows a smaller increase for 
$M_{ij}$. The resulting error in $c_s^2$ for the \second\ order approximation 
with \fourth\ order finite differences is much larger ($37 \%$) than when 
\second\ order 
finite differences are used.   When the \fourth\ order correction 
terms are included, evaluated with \fourth\ order finite differences, the 
correlation coefficients for the $L_{ij}$ and $M_{ij}$ components do not change 
significantly.  In contrast, the normalized square error for the $L_{ij}$ 
obtained with the \fourth\ order approximation and \fourth\ order differences is 
decreased significantly from both the \second\ order case with the \fourth\ order
finite differences and the \fourth\ order case with only \second\ order finite 
differences. The normalized square errors for the $M_{ij}$ calculated using 
\fourth\ order finite differences are increased from both the \second\ order 
approximation with \fourth\ order finite differences, and the \fourth\ order 
approximation with \second\ order finite differences. This shows that use of 
$\Ord(\Delta^4)$ finite differences can reduce error in using the \fourth\ order 
approximation, but overall the results are not as good as when just \second\ 
order finite differences are used to implement the \second\ order approximation.  
% begin Referee C
Since the leading error term in (\ref{eqn:T4D4}) is at higher than fourth order, 
the above analysis of the error terms does not clearly explain the worsening of 
results in going from $T=D=2$ to $T=D=4$. A more intuitive analysis based on the 
roughness of the underlying velocity fields is given in the following section 
(\S~\ref{sec:1.2.4}).
%end

\subsubsection{Effects of smoothness}
\label{sec:1.2.4}
As a check to see under  what conditions the \fourth\ order approximation does 
give better results than the \second\ order approximation, the {\it a priori} 
tests are repeated with velocity fields of varying smoothness.  A 
smoother velocity field allows more accurate calculation of high order 
derivatives, and also yields smaller errors in truncating the Taylor series.  
As a simple way to adjust the smoothness of the velocity field, an additional 
Gaussian filter is applied to the DNS data before performing the {\it a priori} 
tests.  The width of this additional filter, or prefilter, is varied between 
$\Delta$ and $8\Delta$.  This analysis is done here only to illustrate trends and 
is not proposed as a practical method in a simulation where we wish to avoid the 
need for filtering in the first place.

The results obtained with the prefilter width set to $4\Delta$ are summarized in 
the fifth column of Table~\ref{table:results}.  When \second\ order finite 
differences are used, it is seen that the normalized mean square error is 
increased for some components by going from the \second\ order approximation to 
the \fourth\ order approximation.  The relative error in $c_s^2$ is still higher 
for the \fourth\ order approximation. The results obtained by using both 
\fourth\ order finite differences and  prefiltering are shown in the sixth 
column of Table~\ref{table:results}.  In this case, the normalized mean square 
error for the $L_{ij}$ components is decreased, while the results are mixed for 
the $M_{ij}$. The final value of $c_s^2$, as seen in Table~\ref{table:results} in
this case does show an improvement in going from the \second\ order 
approximation to the \fourth\ order approximation.  This trend becomes more 
pronounced as the size of the prefilter is increased further. 
It is important to note that the resulting coefficients should not be compared 
to those obtained without prefiltering, because the velocity field in this case 
is actually different (it has been artificially smoothed out by the prefilter).  

We conclude that only when the velocity field is very smooth does 
the use of the \fourth\ order approximation yield more accurate results.
However, this is not typically the case as LES fields are inherently somewhat 
rough.   This result suggests that under typical circumstances, the \fourth\ order 
approximation will yield poor results when compared to the \second\ order 
approximation.  
   
\subsubsection{Effect of varying $\alpha$}
\label{sec:1.2.5}
To examine how the size of  the test filter affects the accuracy of the 
derivative based method, {\it a priori} tests are done with $\alpha=3$ and $4$,
in addition to the base case of $\alpha=2$ discussed above.  The derivatives in 
these tests are calculated using \second\ order finite differences and keeping 
the truncation to the second-order (i.~e. $D=2$ and $T=2$).  Since the 
derivative based method is based on truncating a Taylor series, one expects that 
the method will give better results for smaller values of $\alpha$, for a given 
$\Delta$.  The correlation coefficients and normalized square errors for
$L_{ij}$ and $M_{ij}$ are shown in the last two columns of 
Table~\ref{table:results}.  For $\alpha=3$, the correlation coefficients 
for $L_{ij}$ and $M_{ij}$ are decreased from their values obtained with 
$\alpha=2$.  The normalized square error of the $L_{ij}$ obtained for the 
\second\ order approximation with $\alpha=3$ are increased dramatically over 
those obtained with $\alpha=2$, especially the off-diagonal terms, which show errors 
$\sim 120 \%$.  For the $M_{ij}$, the error obtained with the \second\ order 
approximation also shows a large increase in error compared to the case 
$\alpha=2$.  The final results for $c_s^2$ are given in Table~\ref{table:results},
and they show that the case  $\alpha=3$  has a $6.41 \%$ relative error.  
Considering the errors in the individual components of $L_{ij}$ and $M_{ij}$,
this result is surprisingly good, but is most likely due to fortuitous 
cancellation of errors.  

As seen in the last column in Table~\ref{table:results},  for $\alpha=4$ the 
correlation coefficients for the $L_{ij}$ are decreased from  the $\alpha=2$ 
and $\alpha=3$ cases.  The correlation coefficients for the $M_{ij}$ are also 
decreased when choosing $\alpha=4$.  The mean square error for both the $L_{ij}$ 
and the $M_{ij}$ is increased significantly in the $\alpha=4$ case.  The error for
 $c_s^2$ obtained with $\alpha=4$ is seen to be very large. Analysis of the 
higher-order case ($T=4$, not shown) yields even larger errors. These {\it a 
priori} tests show that the Taylor series approximation to test filtering yields 
unsatisfactory results for $\alpha \gtrsim 3$ in this isotropic turbulent flow.            

\section{{\em A posteriori} tests}
\label{sec:3}
As a complement to the results of the {\it a priori} tests, {\it a posteriori} 
tests are used to compare the Taylor series based dynamic model to the 
traditional method based on test filtering.  The tests are done by performing 
LES of both forced isotropic turbulence and channel flow.  Since the best 
results in the {\it a priori} tests were obtained by truncating the Taylor 
series expansions at second order, all {\it a posteriori} tests are done using
$T=2$ and $D=2$.  

\subsection{LES of Forced Isotropic Turbulence}
\label{sec:3.1}
To test the derivative-based method in isotropic turbulence, the pseudo-spectral 
DNS code is modified to perform LES on a $64^3$ grid.   This includes adding a 
$3/2$-rule dealiasing procedure, which is applied only to the nonlinear term in 
the filtered Navier-Stokes equations ($\widetilde{\vct{u}}\times\widetilde{\vct{\omega}}$) and not the 
SGS stress term. This is done because the nonlinear term is known to be exact, 
while the SGS term is already an approximation.  In addition, the type of 
nonlinearity contained in the SGS stress term is not completely removed by 
standard dealiasing procedures such as the $3/2$-rule or random phase shifts.
Simulations are forced with a constant energy injection rate \citep[as in][]{Ceruttietal98} of $\varepsilon = 0.0007$, the molecular viscosity is $\nu=10^{-6}$, and the
domain is of length $L=2\pi$.  At statistical steady state, the Taylor-scale Reynolds
number of the flow is $R_\lambda \approx 2100$.  
Simulations using both exact test filtering and the Taylor series based method 
are performed.  As with the {\it a priori} tests, a Gaussian test filter is 
used at scale $2\Delta$ (corresponding to $\alpha=2$).   The exact test 
filtering is done in Fourier space, while the Taylor series method uses second 
order finite differences to calculate all of the derivatives associated with the 
dynamic procedure.   The initial condition is a random field with a $k^{-5/3}$ 
spectrum and random phases. Volume averaging is used in this spatially 
homogeneous flow.  

\begin{figure*}[p]
\centerline{%
\begin{tabular}{c@{\hspace{6pc}}c}
\includegraphics[width=2.5in]{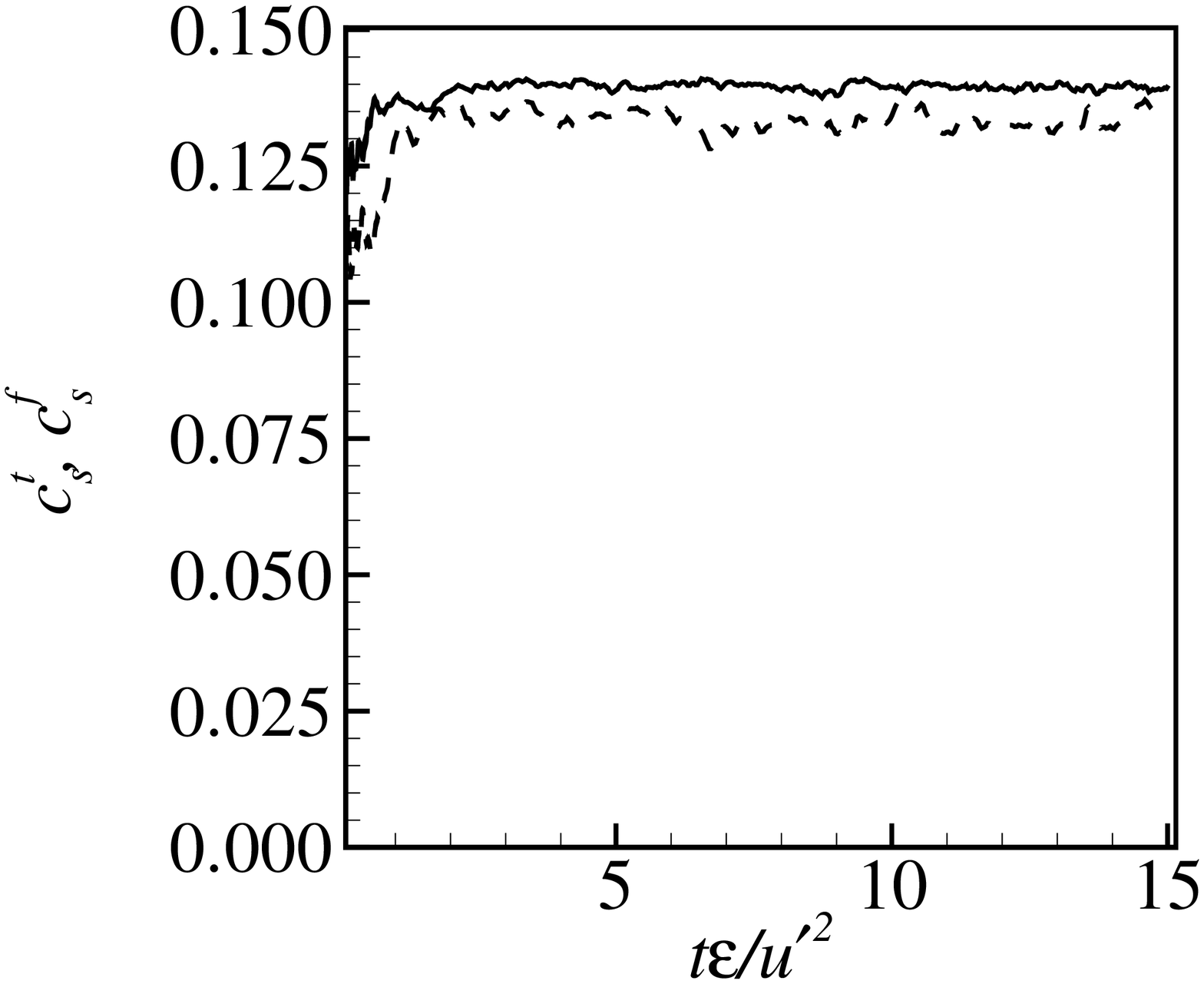} &
\includegraphics[width=2.5in]{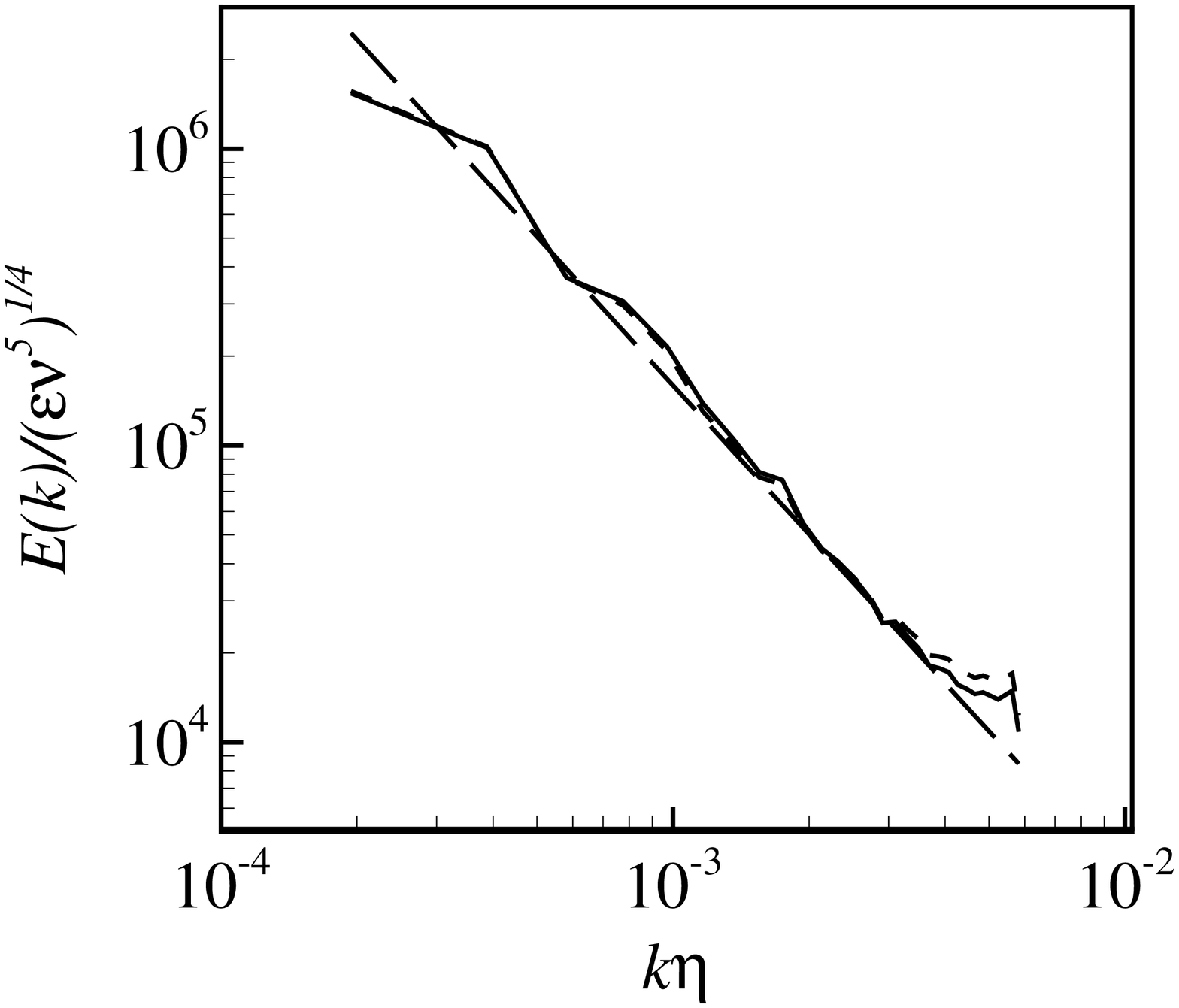} \\
(a) & (b)
\end{tabular}}
%\centercap
\caption{Results from {\it a posteriori} tests in LES of forced isotropic turbulence on a $64^3$ domain with $R_\lambda \approx 2100$. (a) time evolution of dynamic Smagorinsky coefficients. Solid line: derivative-based, dashed line: test-filter based. (b) radial energy spectra. Solid line: derivative-based, dashed line: test-filter based dynamic model, long dashed line: normalized Kolmogorov spectrum $c_K(k\eta)^{-5/3}$ with $c_K=1.6$.}
\label{fig:LESpost}
\end{figure*}

Figure~\ref{fig:LESpost}(a) shows the time evolution of $c_s$ 
for exact test filtering 
and the derivative based approach.  Although the derivative based method 
consistently gives a value of $c_s$ roughly $5$ \% higher than the exact method
(in contrast to the a-priori tests),  both methods relax from the initial value 
($c_s(0)=0$, as expected for Gaussian fields) to their quasi-steady value at 
about the same rate.   Most importantly, the calculation done using the 
derivative based method is seen to be stable.  The quasi-steady values for both 
methods lie near $c_s \approx 0.13$ and are lower than the value of 
$c_s \approx 0.16$ often observed in isotropic turbulence.  This may be a result 
of using a Gaussian filter.    As shown in Figure~\ref{fig:LESpost}(b), the radial 
energy spectra obtained from the two methods are essentially the same, only 
showing a slight difference at high wavenumber due to the slightly different 
values of $c_s$.   Both cases agree well with the Kolmogorov $-5/3$ prediction, 
with $c_K=1.6$.  

\subsection{Application to channel flow}
\label{sec:3.2}
To examine the effects of anisotropy and the presence of walls on the 
derivative-based procedure, moderate Reynolds number channel flow simulations 
are performed with a LES version of the DNS code NTMIX3D \citep{Stoessel}. 
In this code for fully compressible flow, the governing equations are integrated 
using an explicit third order  Runge Kutta time advancement and a sixth order 
compact finite difference  scheme \citep{Lele}.  No slip isothermal boundary 
conditions  are used at the top and bottom boundaries, while periodic boundary 
conditions  are used in the streamwise ($x$) and spanwise ($z$) directions 
(which are homogeneous directions). A driving source term compensates the shear 
stress at the walls  which allows to run the simulations using periodicity in 
the $x$ direction. A computational domain of size $L_x=2\pi$, $L_y=2$ ($=2\delta$), 
and $L_z=0.908$ in the three directions is used. The $x$, $y$, and $z$ coordinates 
belong to the intervals  
$0\leq x \leq L_x$, $-L_y/2 \leq y \leq L_y/2$,  and $0 \leq z \leq L_z$.

The shear Reynolds number (based on friction velocity $u_{\tau}$ and channel 
half width $\delta$) is $Re_{\tau}=180$, which corresponds to a convective 
Reynolds number $Re_c=3300$ (based on channel half width and axial velocity). 
To allow a reasonable time step within the restrictions of the acoustic CFL 
condition, the mean centerline Mach number $M=0.2$ has been  chosen in the low 
subsonic domain. 
% begin Referee B
The trace of the sub-grid scale stress tensor can be rewritten as 
$\tau_{kk}=\gamma M^2_{SGS}\overline{p}$. For the small Mach number 
considered in the present simulations, the sub-grid Mach number 
$M_{SGS}$ is expected to be small. Consequently we simply neglect 
$\tau_{kk}$. Also, SGS fluxes in the total energy equation were treated
using a fixed (non-dynamic) value $\mathrm{Pr}_t=0.6$. Because of the low Mach 
number, the linkage to the momentum field was totally negligible.
% end
The simulation parameters have been fixed to allow 
quantitative comparisons with the direct numerical simulation data of 
\citet{kimetal}. Two LES runs with the dynamical Smagorinsky model are 
performed:  one with the classical filtering procedure and the other with the 
Taylor series, derivative-based method. 
 
For both LES cases the computational grid contains 
$17 \times 61 \times 16$ points in the $x,y,z$ directions (or $x_1,x_2,x_3$). 
The grid is uniform in the streamwise and spanwise directions and the 
corresponding resolution was $\Delta_x^+ \approx 66$ and 
$\Delta_z^+ \approx 10$. Following \citet{gamet1}, we use a non-uniform mesh 
in the wall normal direction based on the distribution: 
$y_i=\frac{L_y}{2}{\rm \tanh}(K\eta_i)$ with $K={\rm \atanh}(\frac{1}{C})$ 
and $-1 \leq \eta_i = 2 \frac{i-1}{N_y} - 1 \leq +1$.
The constant $C$ is such that $\Delta_{y}^+\approx 2$ at the wall 
and $\Delta_{y}^+\approx 10$ near the centerline. 
For comparison, minimal channel DNS was performed using $N_x=34$, $N_y=121$ 
and $N_z=32$.  
LES are started from filtering a fully turbulent DNS field. The statistics were 
accumulated over time $tu_{\tau}/\delta=5.4$.

As usual in channel flow simulation with the dynamic procedure, we apply the test 
filter operation only in the streamwise and spanwise directions.  Because of the 
non uniform mesh in the cross direction, we choose to use for the test filter 
width: $\ol{\Delta}=\alpha\left(\Delta_x \Delta_z\right)^{1/2}$, with $\alpha=2$. 
For the LES of the filter-based reference case we apply a box filter, using a 
trapezoidal rule for the integration of the convolution operation. 
For the derivative-based method, the second order approximation ($T=2$, $D=2$) 
has been chosen based on the results presented in section \ref{sec:2.2}.
% begin Referee A
Here the equivalence at second order between the Gaussian and box filters 
mentioned in \S 2 is used.
% end 
The final expression for $\left( c_s^t \right)^2$ is given in 
(\ref{eqn:c_s Taylor}), with planar Laplacians ($k=1,3$ and $m=1,3$) 
and planar averaging, as discussed in \S~\ref{sec:1}.

\subsubsection{{\it A priori} results based on initial condition for LES}
\label{sec:3.2.1}
\begin{figure*}[p]
\centerline{\hbox{
\includegraphics[width=8cm]{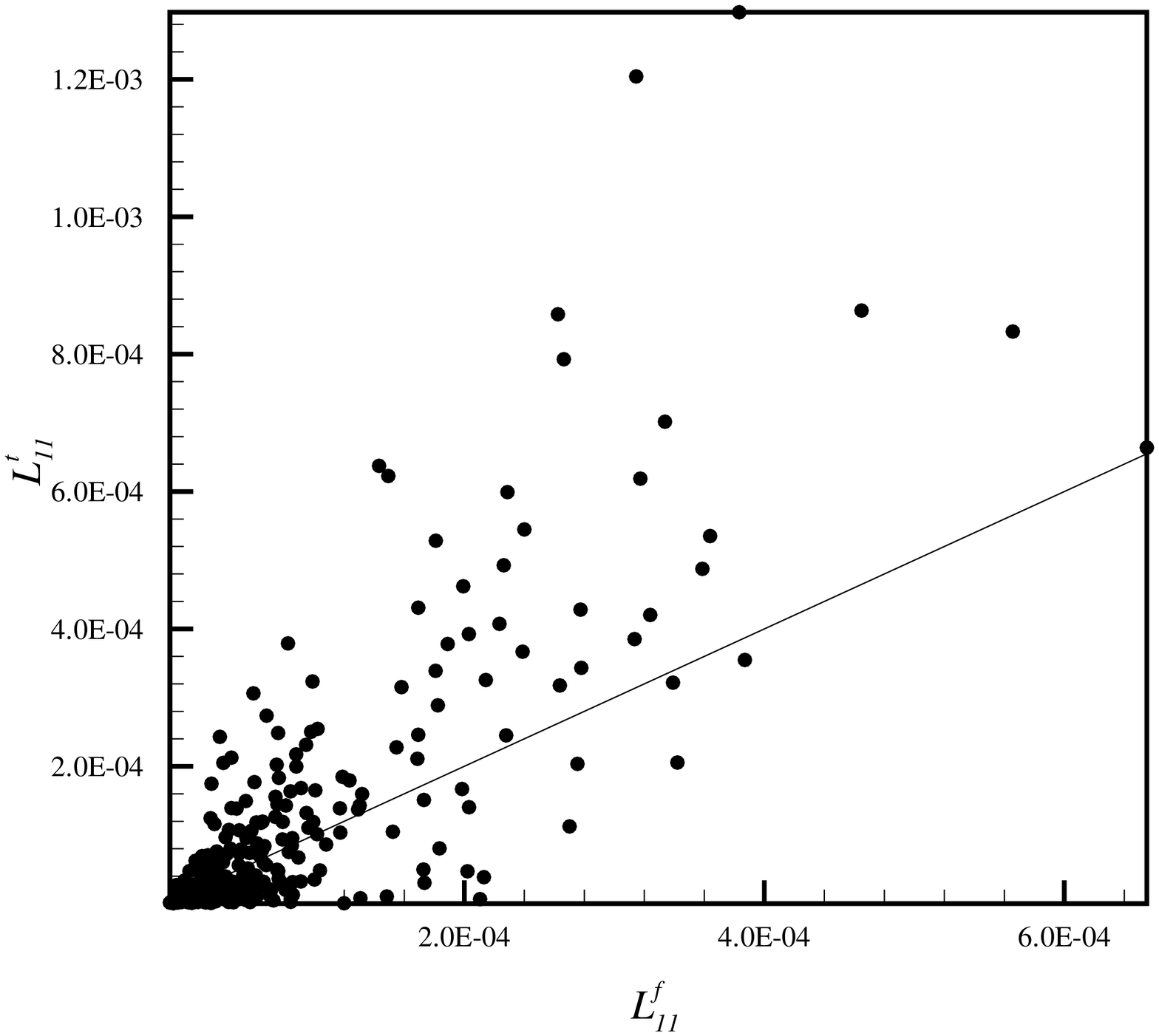}
\includegraphics[width=8cm]{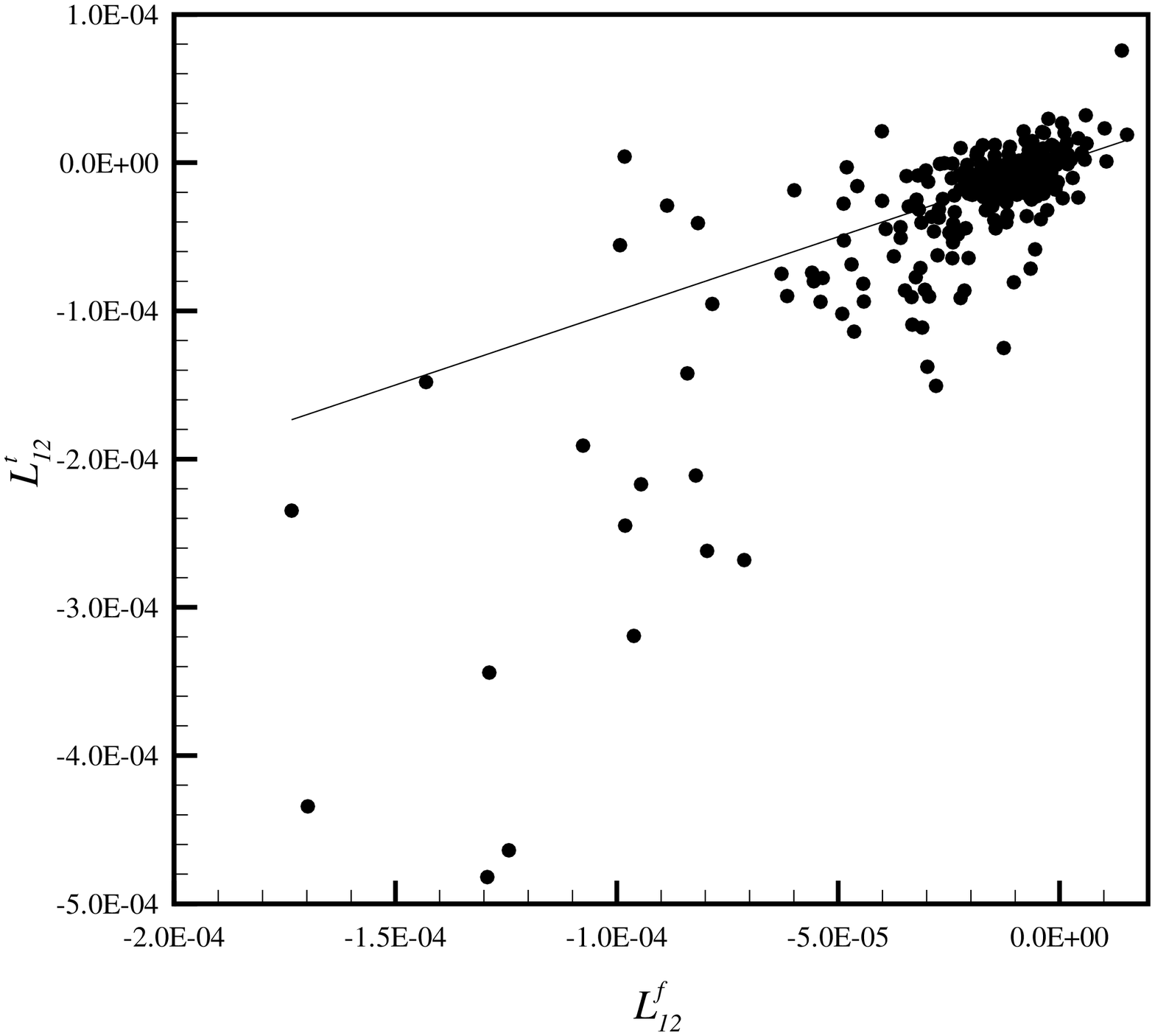}
}}
\centerline{\hbox{
\includegraphics[width=8cm]{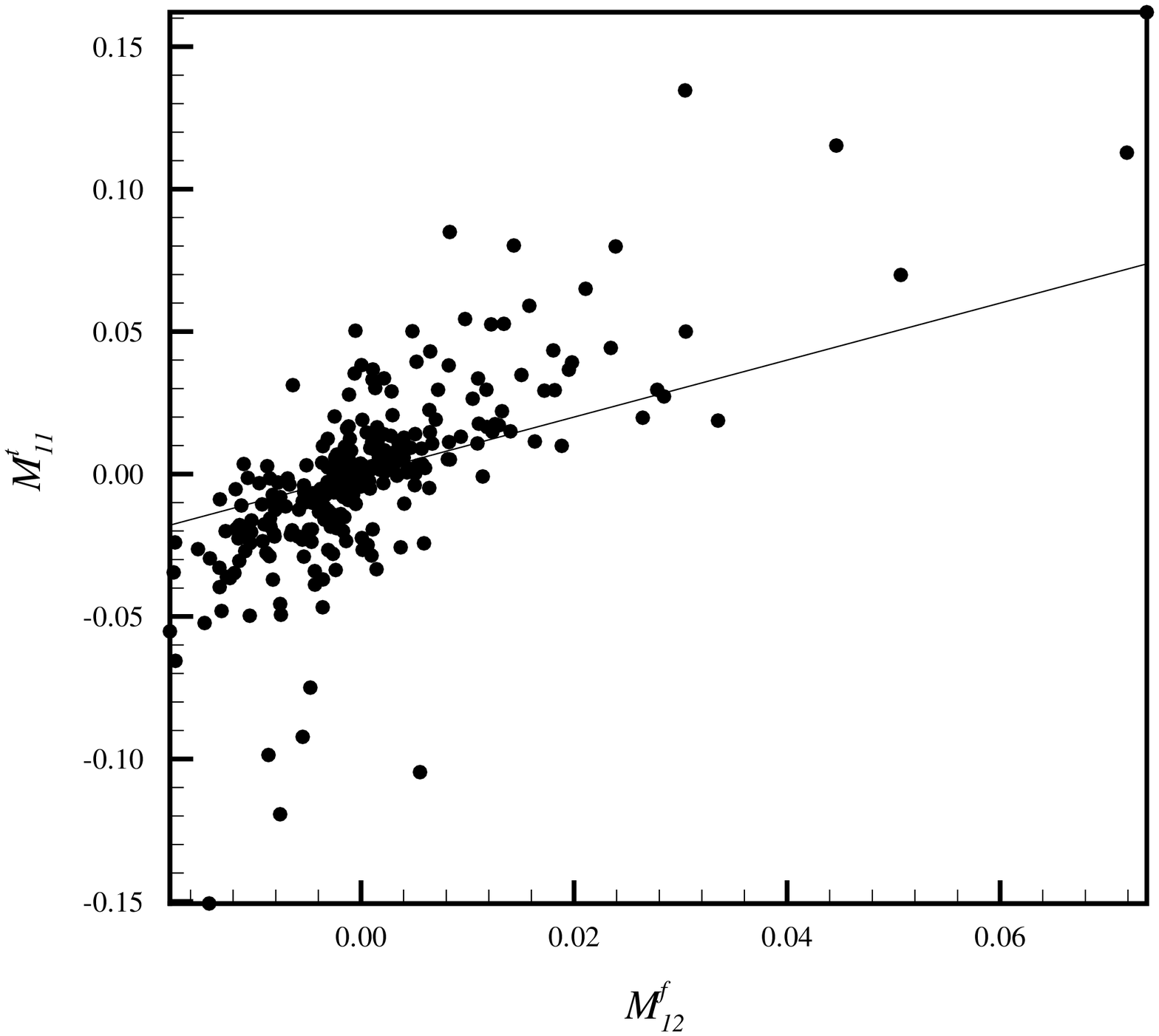}
\includegraphics[width=8cm]{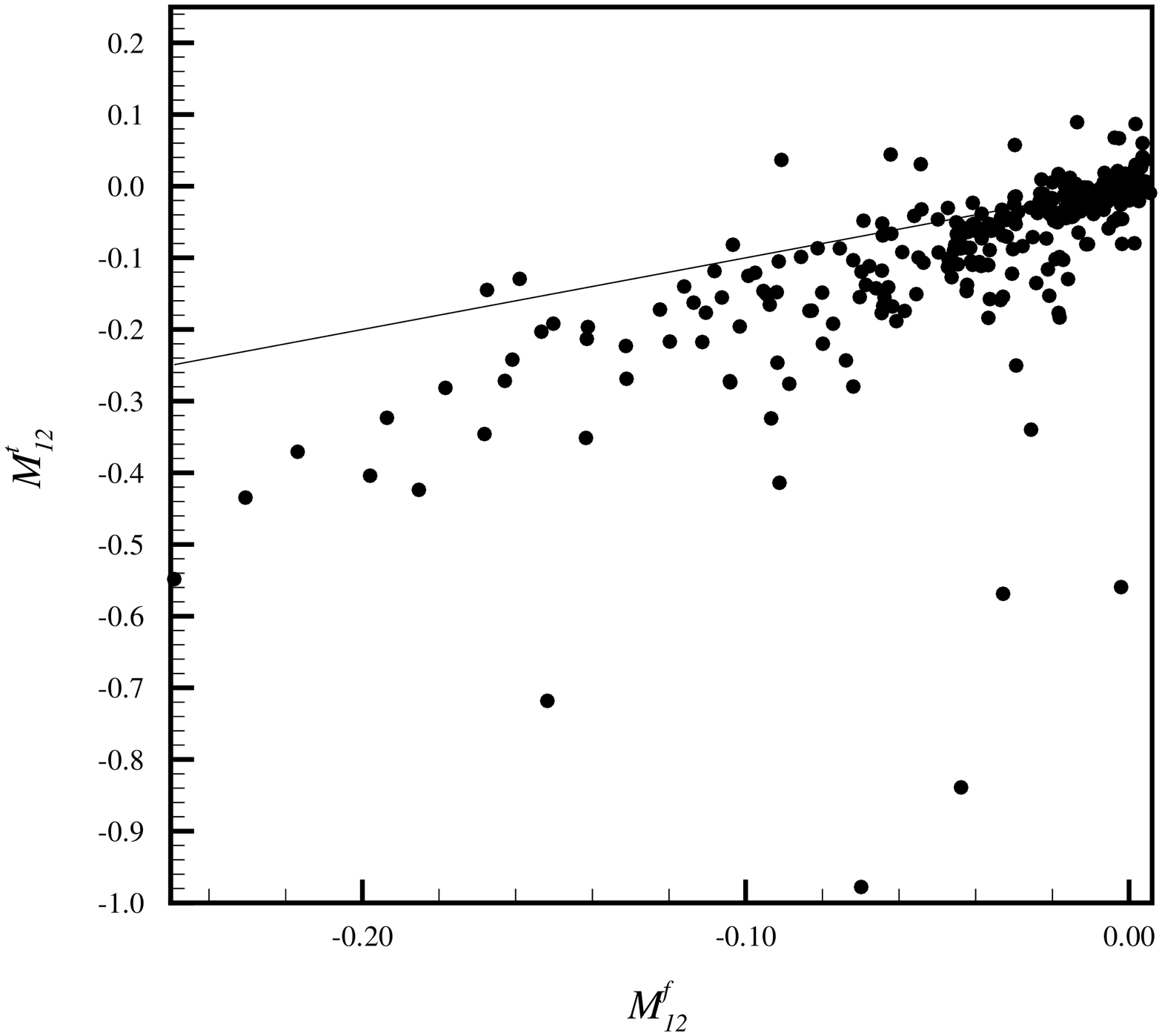}
}}
%%\centercap
\caption{Scatter plots of the $L_{11}$, $L_{12}$, $M_{11}$, $M_{12}$ components in the $x-z$ plane at $y^{+}=43$ from filtered DNS of channel flow, comparing local values obtained from the derivative-based approach with the classical filter-based values.  
}
\label{fig:scatterplotL11L12M11M12}
\end{figure*}

\begin{figure*}[p]
\centerline{\hbox{
\includegraphics[width=8cm]{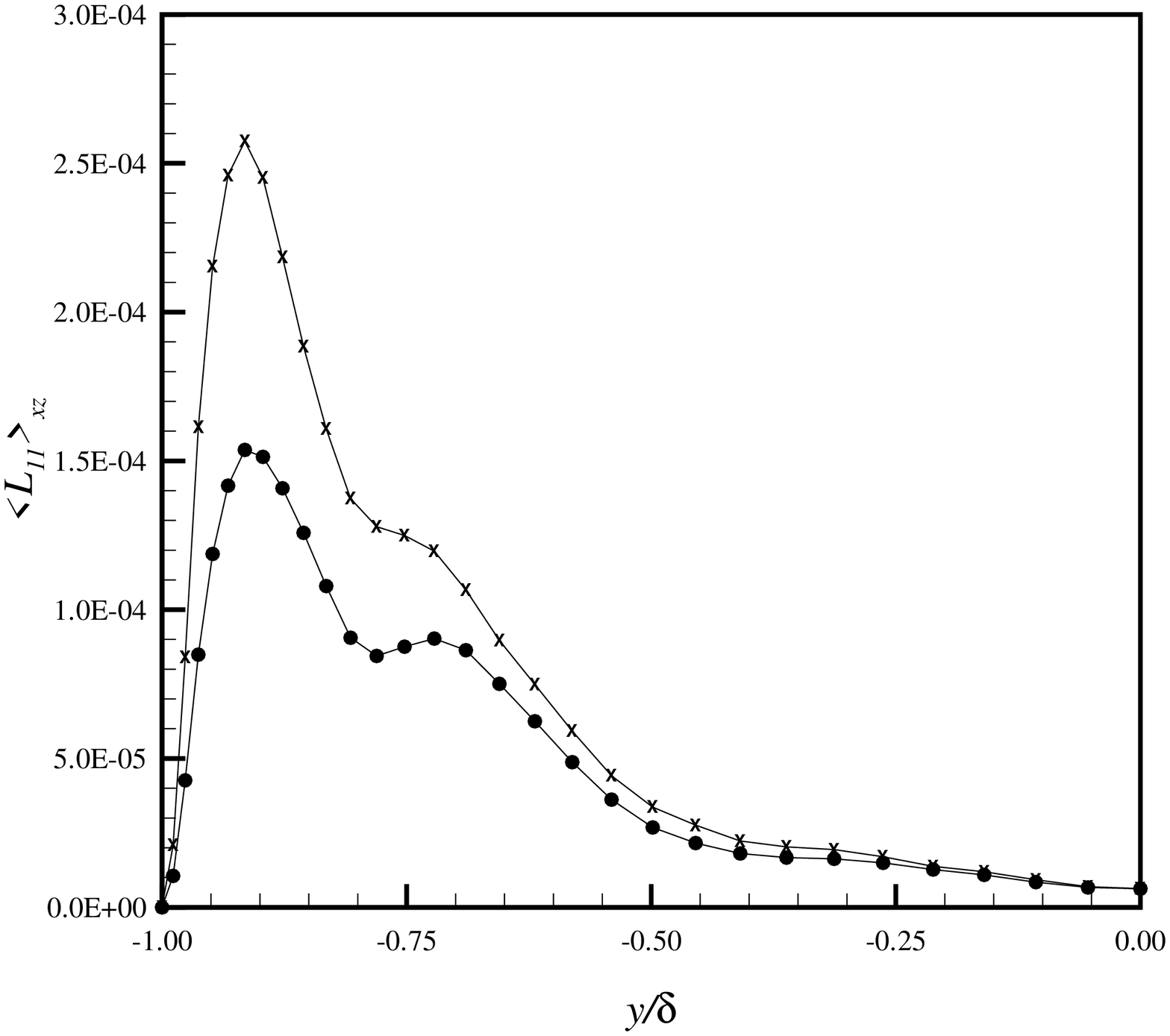}
\includegraphics[width=8cm]{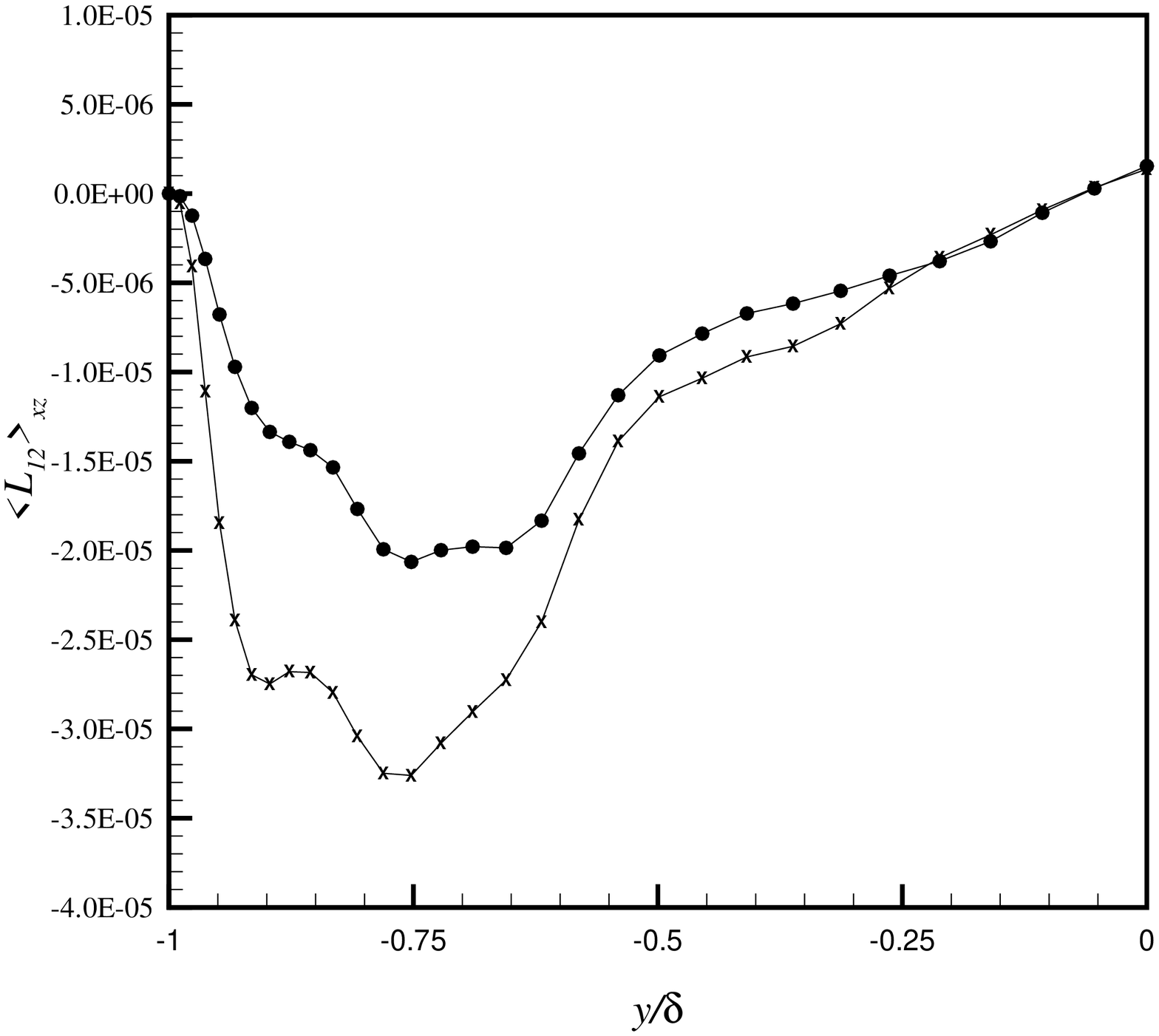}
}}
\centerline{\hbox{
\includegraphics[width=8cm]{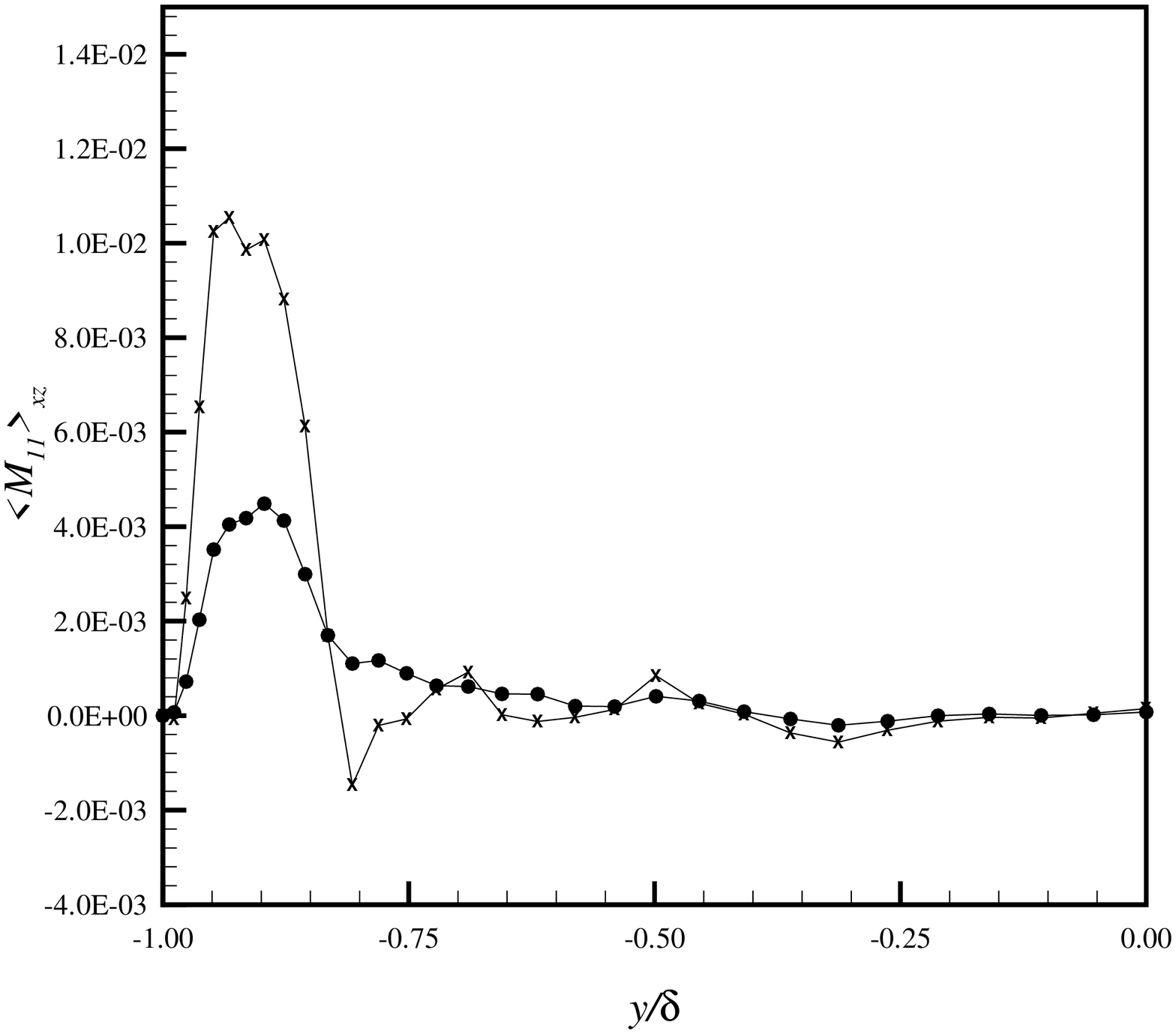}
\includegraphics[width=8cm]{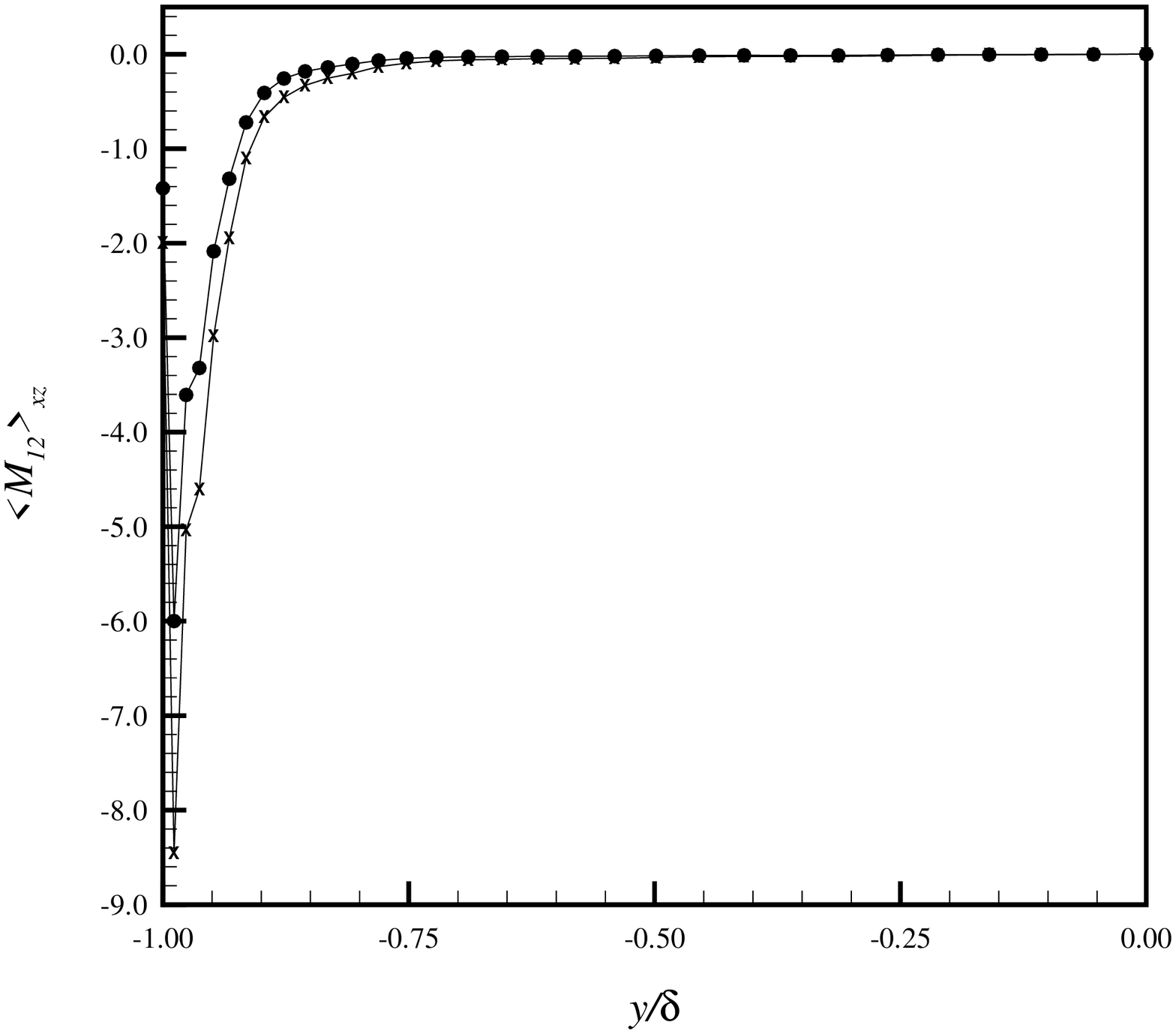}
}}
%%\centercap
\caption{Comparisons of the mean 
components $L_{11}$, $L_{12}$, $M_{11}$, $M_{12}$ averaged in the $x$-$z$ planes across the wall normal direction. $\times$ derivative-based method, $\bullet$ classical filter-based method.}
\label{fig:L11L12M11M12}
\end{figure*}

As a first step we perform {\it a priori} tests at the initial time of the LES
simulations, which are filtered DNS fields. Scatter plots comparing the 
$L_{11}$, $L_{12}$, $M_{11}$, $M_{12}$ components  obtained by the Taylor series 
and filtering based methods, in the $x$-$z$ plane at $y^{+}=43$, are shown in 
Figure~\ref{fig:scatterplotL11L12M11M12}.  The agreement between real filtered and 
estimated local values is fair. The general trends are reproduced with 
correlation coefficients of order 0.7. Specifically 
$\rho(L^t_{11},L^f_{11})=0.79$, $\rho(L^t_{12},L^f_{12})=0.81$, 
$\rho(M^t_{11},M^f_{11})=0.74$, and $\rho(M^t_{12},L^f_{12})=0.64$. Even
if the local behavior of the typical components of $L_{ij}$ and $M_{ij}$ is 
not highly accurate, the global behavior of these components is estimated 
reasonably well, as demonstrated by Figure~\ref{fig:L11L12M11M12}, showing 
the same tensor elements averaged in the $x$-$z$ planes, as a function of wall normal 
direction.  We notice that despite an overestimation of the magnitude of each 
component, their general shape is correctly predicted by the derivative-based 
approximation.  

\begin{figure}[p]
\centerline{\hbox{
\includegraphics[width=8cm]{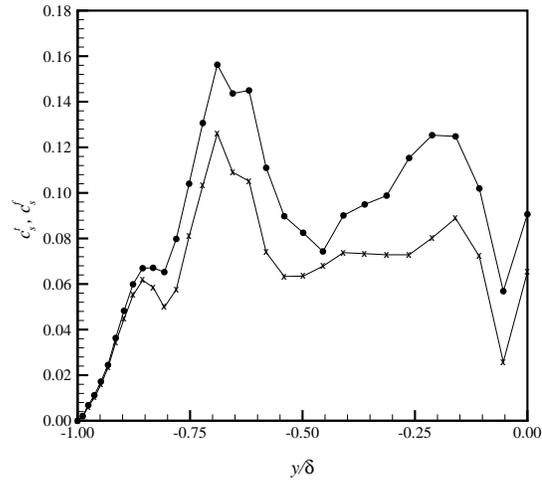}
}}
%\centercap
\caption{Dynamic Smagorinsky coefficients, obtained {\it a priori} from filtered DNS of minimal channel flow. $\times$ derivative-based method, $\bullet$ classical filter-based method.}
\label{fig:apriorismagorinskycs}
\end{figure}

Figure~\ref{fig:apriorismagorinskycs} shows the instantaneous Smagorinsky 
coefficients obtained by the classical filtering procedure and by the present 
derivative-based method.  The {\it a priori} test shows that the Smagorinsky 
coefficient obtained by the Taylor series expansion approach is slightly smaller than 
the classical filter-based results.  However, the agreement is sufficiently good 
in the context of a practical scheme, especially the reduction to zero in the 
well-resolved near-wall region.  As we will see in the next section, 
{\it a posteriori} tests give improved results.  
 
\subsubsection{{\it A posteriori} results of LES}
\label{sec:3.2.2}
\begin{figure*}[p]
\centerline{\hbox{
\begin{tabular}{c@{\hspace{6pc}}c}
\includegraphics[width=8cm]{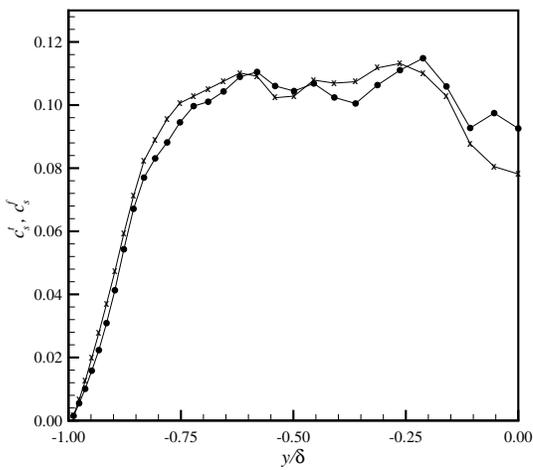} &
\includegraphics[width=8cm]{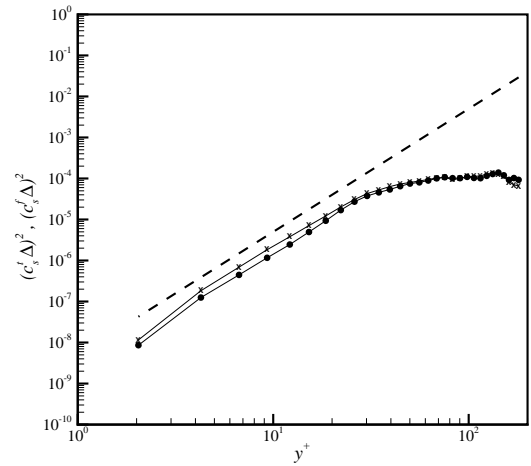}\\
(a) & (b)
\end{tabular}
}}
%\centercap
\caption{Comparisons of the Smagorinsky coefficients obtained in {\it a posteriori} tests by the filter-based and derivative-based methods in the global (a) and wall (b) coordinates.  Same symbols as in Figure~\ref{fig:apriorismagorinskycs}. Dashed line is the expected $c_s^2 \Delta^2 \sim (y^+)^3$ behavior.}
\label{fig:csandcsdelta2}
\end{figure*}

All the results presented in this section are time-averaged over 55 realizations 
spaced by a time interval of $0.1 \delta/u_{\tau}$.  In 
Figure~\ref{fig:csandcsdelta2}(a), we observe that the Smagorinsky coefficient is 
very well predicted by the derivative-based method, except some deviation near the 
centerline. As shown in Figure~\ref{fig:csandcsdelta2}(b) the derivative-based 
approach yields the expected $y^{+3}$ behavior in the vicinity of the wall. 

\begin{figure*}[p]
\centerline{\hbox{
\includegraphics[width=8cm]{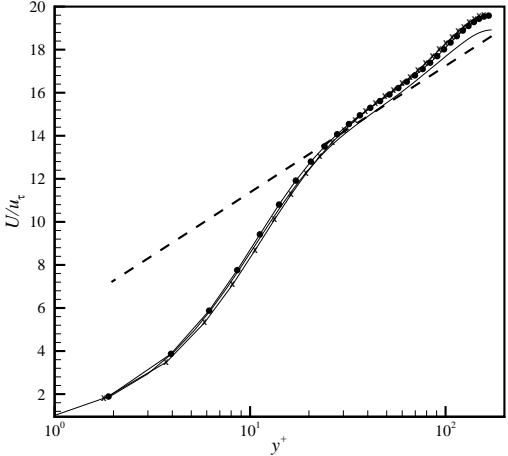}
}}
%\centercap
\caption{Mean Velocity U in wall coordinates: $-$ DNS, $\times$ derivative-based method, $\bullet$ classical filter-based dynamic model. Dashed line: $U/u_\tau = 2.5\ln(y^+)+5.5$.}
\label{fig:Umean}
\end{figure*}

In Figure~\ref{fig:Umean}, we observe a fairly good agreement in streamwise mean 
velocity $U$ between the results of the Taylor series expansion 
method and both the results of the classical filter-based dynamic model and the DNS results. 
Despite a slight underestimation in the buffer region, the proposed 
approach leads to quite accurate results. 

\begin{figure*}[p]
\centerline{\hbox{
\begin{tabular}{c@{\hspace{6pc}}c}
\includegraphics[width=8cm]{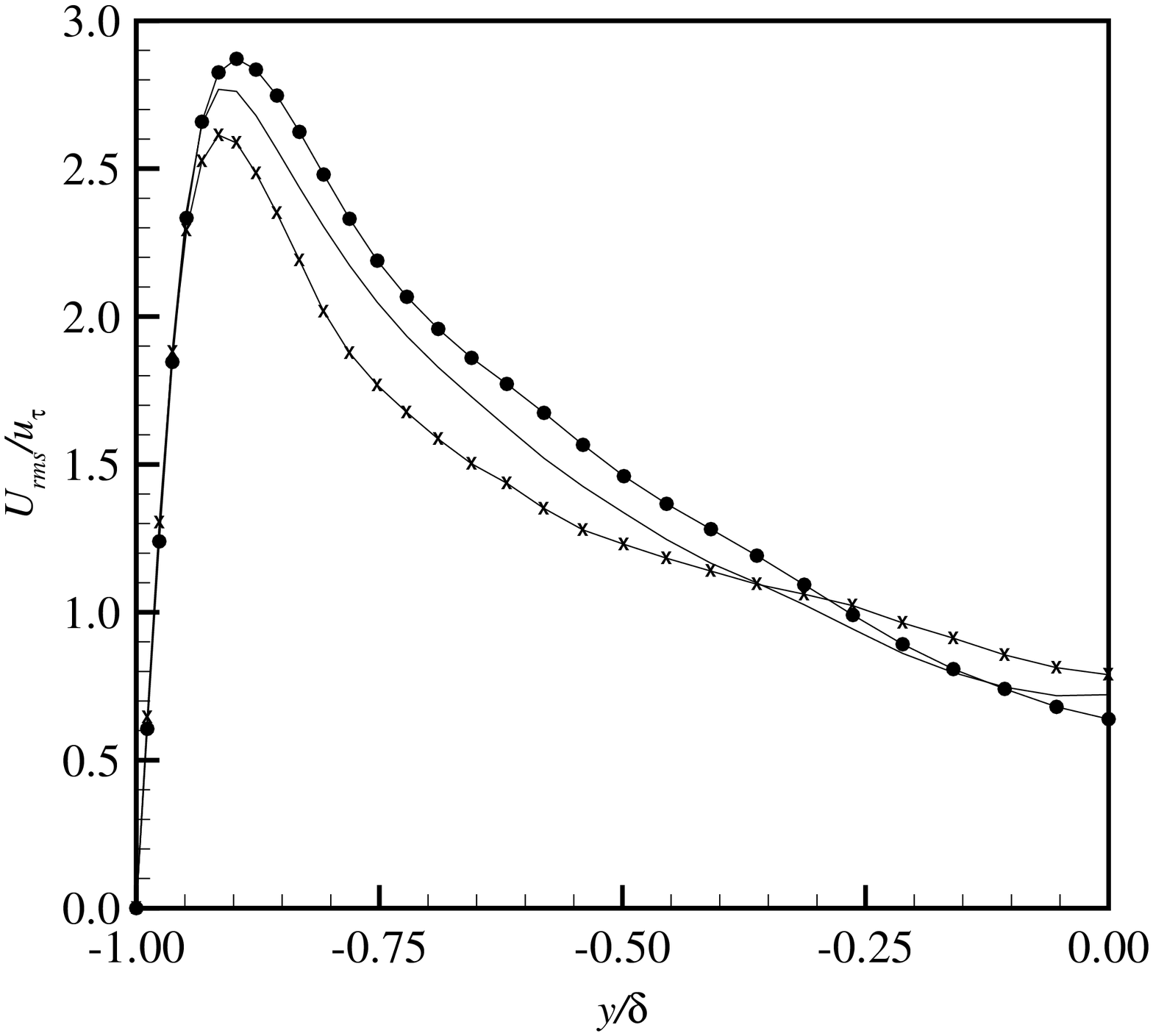}&
\includegraphics[width=8cm]{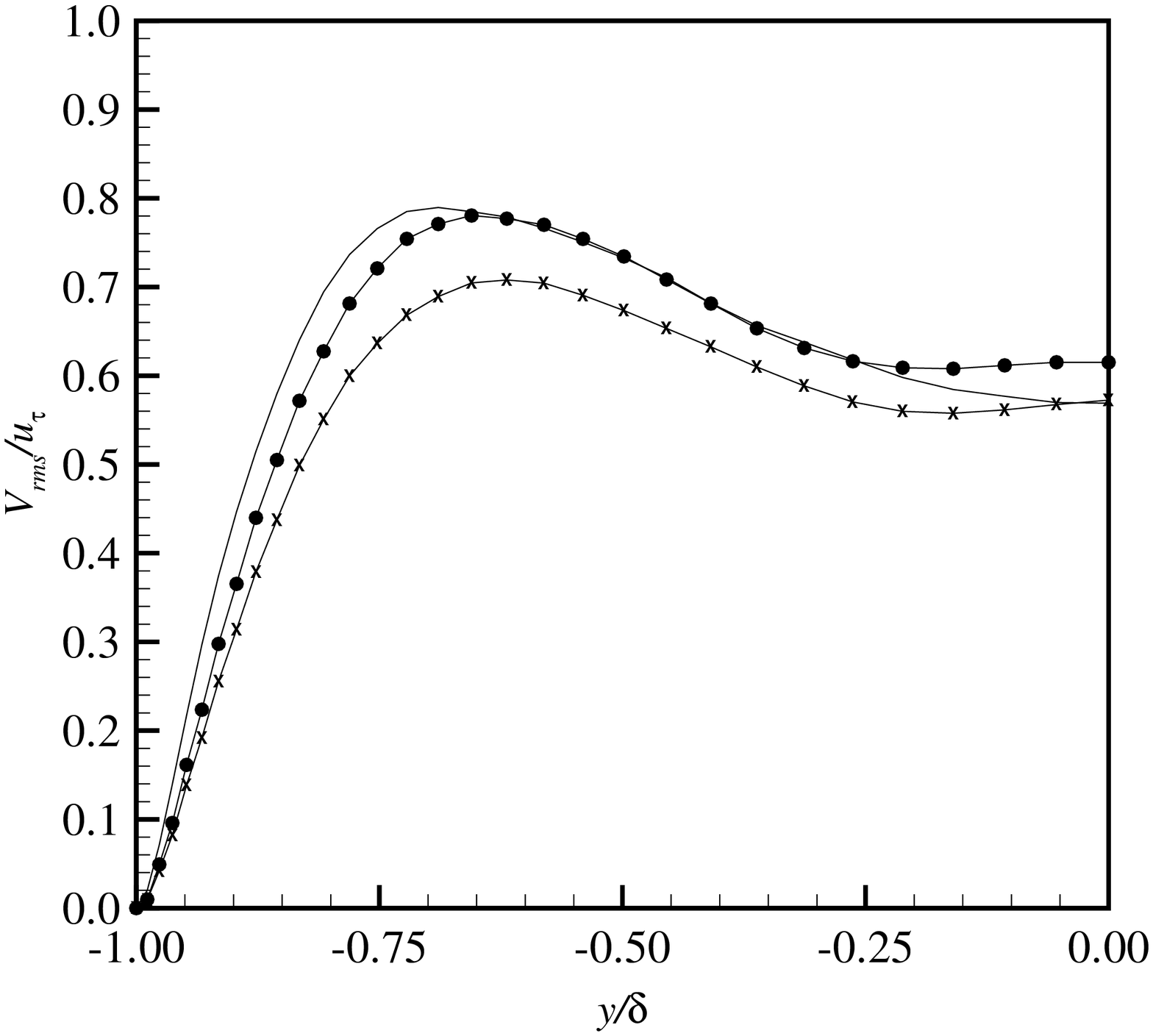}\\
(a) & (b)
\end{tabular}
}}
\centerline{\hbox{
\begin{tabular}{c@{\hspace{6pc}}c}
\includegraphics[width=8cm]{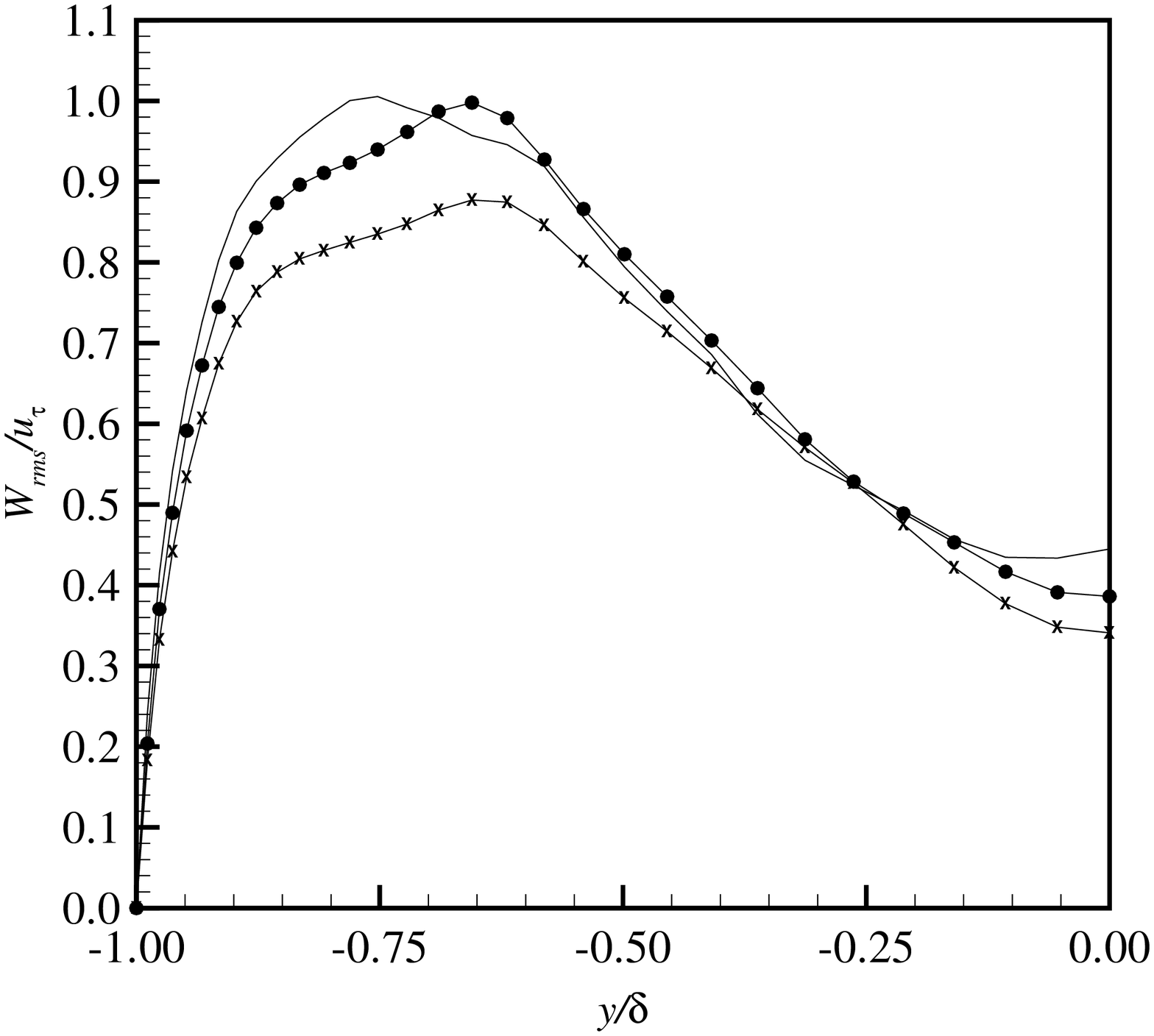}&
\includegraphics[width=8cm]{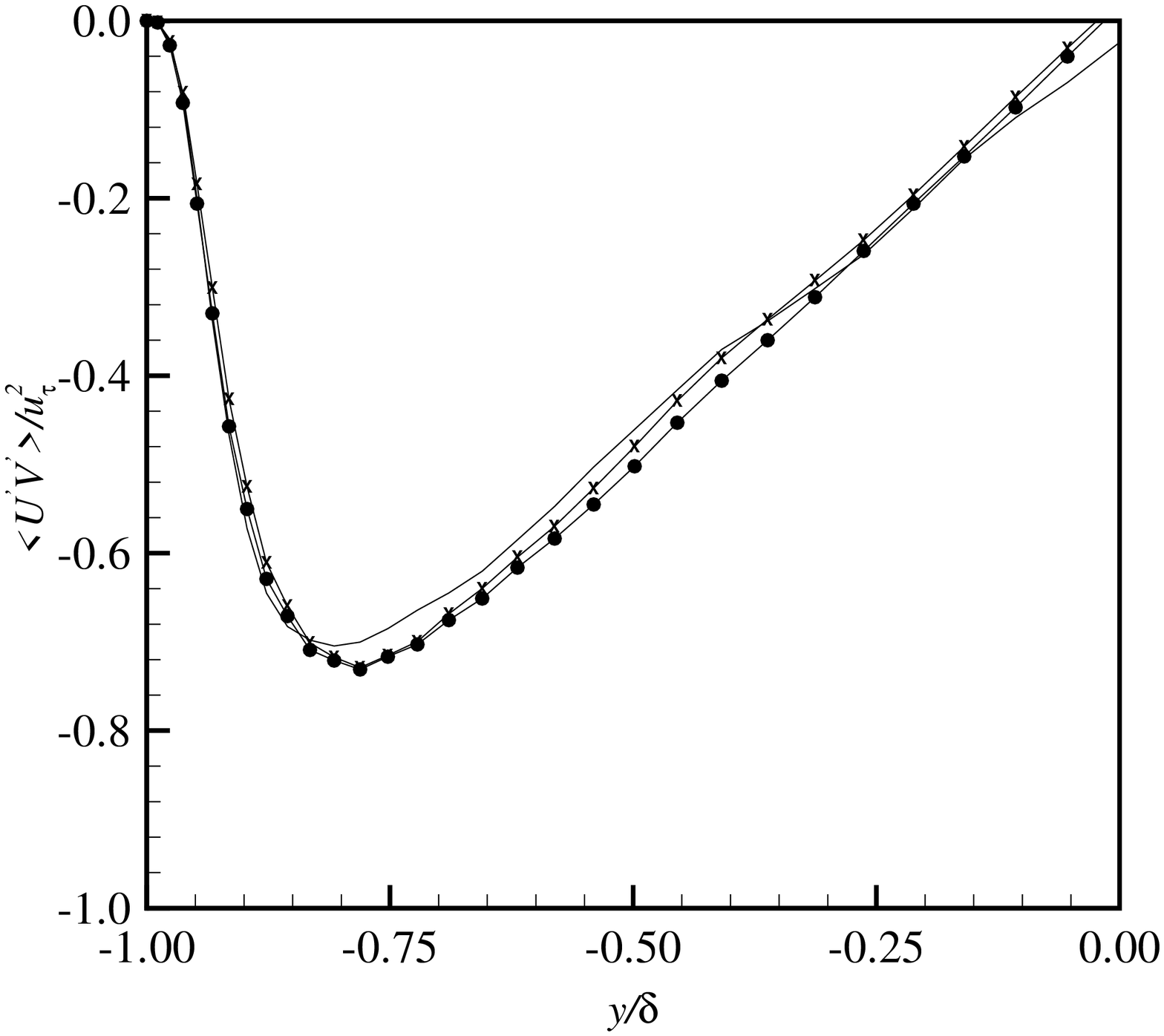}\\
(c) & (d)
\end{tabular}
}}
%\centercap
\caption{(a)-(c) Comparisons between rms velocities from LES and filtered DNS velocity rms. (d) Comparison between Reynolds shear stresses $\langle \widetilde{u}^{\prime}\widetilde{v}^{\prime}\rangle$ from LES and filtered DNS. Same symbols as in Figure~\ref{fig:Umean}.}
\label{fig:rmsandUpVp}
\end{figure*}

In Figure~\ref{fig:rmsandUpVp} are shown the rms intensities of the large-scale  
(resolved) turbulent velocity fluctuations.  These rms quantities are compared 
with the rms velocities from the DNS filtered at scale $\Delta$ using a box 
filter. The agreement between both LES results and the filtered DNS data is quite
good.  For the derivative based method, the streamwise rms velocity is slightly 
overestimated by both LES approaches.  The wall-normal and transverse  rms 
velocities are underestimated slightly more by the derivative-based method than 
the filter-based method.  In Figure~\ref{fig:rmsandUpVp}(d), we compare the Reynolds 
shear stress of the resolved velocity, 
$\langle \widetilde{u}^{\prime}\widetilde{v}^{\prime}\rangle$. The proposed
approach yields very good agreement with the filter-based dynamic model and the 
reference filtered DNS data.

\section{Conclusions}
\label{sec:4}
An implementation of the dynamic Smagorinsky model that uses Taylor series 
expansions to avoid explicit test-filter operations in LES has been developed. 
The proposed method has been subjected to {\it a priori} and {\it a posteriori} tests.
These tests were performed using structured grids, to check the performance of 
the method in idealized, well controlled reference cases.  

The {\it a priori} results obtained by truncating the Taylor series at 
\second\ order and at \fourth\ order have been compared with results obtained by 
test filtering.  It was found that for LES of 
isotropic turbulence, it is possible to obtain values of $c_s^2$ accurate to 
$\sim 5$ \% by using the \second\ order approximation.  However, results are not 
improved by using the \fourth\ order approximation because of the errors 
associated with evaluating high-order derivatives on inherently rough LES fields.  
From these {\it a priori} tests, it is concluded that the derivative based method 
should be implemented using the \second\ order approximations, with 
$\alpha =2$. {\it A posteriori} tests of forced isotropic turbulence show that 
the derivative-based method yields stable results, and values of $c_s^2$ to 
within $\sim 10 \%$ of those obtained by explicit test filtering.   

Applications to a minimal channel configuration show that the observed 
differences between the tensor elements in the  filter-based and Taylor series 
based approaches are larger than those in isotropic turbulence.  This could be 
due to the strong anisotropy of the test filter as well as of the 
turbulence itself. However the Smagorinsky coefficient is correctly estimated 
by the proposed approach, especially the $y^{+3}$ behavior in the  vicinity of 
the wall. Moreover first and second order statistics, which are the most 
important ones for practical engineering calculations, are correctly predicted 
when using the derivative-based dynamic model. 

Strictly speaking, these conclusions are applicable only to the fairly simple 
flow configurations considered in this work. However, the results suggest that 
applications of the Taylor series based dynamic model to complex-geometry flows 
on unstructured grids is a promising direction.  
Taken together with the results of \citet{Jansen99} on the alternative generalized 
box-filter approach, the present results show that applications of
the dynamic model to LES of complex-geometry turbulent flows are 
feasible.

\newpage
%figures

\newpage
%table
\begin{table}[h]
\caption{Correlation coefficients, normalized square error, values of $c_s^2$, 
and relative error in $c_s^2$ obtained from {\it a priori} tests in forced 
isotropic turbulence.} 
\label{table:results}
\begin{center}
\begin{tabular}{|p{2.5cm}|cccccccc|} 
\hline
\hspace{0.8cm}{$\alpha$} &  \multicolumn{6}{|c|}{2} & \multicolumn{1}{|c|}{3} & \multicolumn{1}{|c|}{4}  \\ \hline

prefiltering &  \multicolumn{4}{|c|}{none}  & \multicolumn{2}{|c|}{$4\Delta$} & \multicolumn{2}{|c|}{none} \\ \hline

order of finite differences ($D=$~)  & \multicolumn{2}{|c|}{\raisebox{-1.5ex}[0pt]{2}} & \multicolumn{2}{|c|}{\raisebox{-1.5ex}[0pt]{4}} & \multicolumn{2}{|c|}{\raisebox{-1.5ex}[0pt]{4}}
& \multicolumn{2}{|c|}{\raisebox{-1.5ex}[0pt]{2}} \\ \hline

order of expansion ($T=$~) & \multicolumn{1}{|c|}{\raisebox{-1.5ex}[0pt]{2}} & \multicolumn{1}{|c|}{\raisebox{-1.5ex}[0pt]{4}} & \multicolumn{1}{|c|}{\raisebox{-1.5ex}[0pt]{2}} & \multicolumn{1}{|c|}{\raisebox{-1.5ex}[0pt]{4}} & \multicolumn{1}{|c|}{\raisebox{-1.5ex}[0pt]{2}} &  \multicolumn{1}{|c|}{\raisebox{-1.5ex}[0pt]{4}} & \multicolumn{2}{|c|}{\raisebox{-1.5ex}[0pt]{2}} \\ \hline
 
%\hspace{0.5cm}
$\cor{L}{11}$ & 0.955 & 0.978 & 0.946 & 0.981 & 0.993    & 1.000    & 0.889 & 0.791 \\
%\hspace{0.5cm}
$\cor{L}{12}$ & 0.947 & 0.976 & 0.932 & 0.977 & 0.993    & 1.000    & 0.863 & 0.748 \\
%\hspace{0.4cm}
$\cor{M}{11}$ & 0.996 & 0.990 & 0.995 & 0.986 & 1.000    & 1.000    & 0.832 & 0.380 \\
%\hspace{0.4cm}
$\cor{M}{12}$ & 0.998 & 0.993 & 0.996 & 0.989 & 1.000    & 1.000    & 0.868 & 0.415 \\[1.5ex]
%\hspace{0.5cm}
$\Err{L}{11}$ & 0.056 & 0.070 & 0.187 & 0.029 & 0.023    & 5.78E-4  & 0.369 & 1.301 \\
%\hspace{0.5cm}
$\Err{L}{12}$ & 0.158 & 0.078 & 0.514 & 0.050 & 0.048    & 9.13E-4  & 1.264 & 4.556 \\
%\hspace{0.4cm}
$\Err{M}{11}$ & 0.009 & 0.031 & 0.014 & 0.047 & 1.08E-4  & 3.03E-4  & 0.363 & 10.315 \\
%\hspace{0.4cm}
$\Err{M}{12}$ & 0.006 & 0.025 & 0.009 & 0.040 & 1.02E-4  & 4.51E-4  & 0.268 & 7.845 \\[1.5ex]
%\hspace{0.9cm}
$(c_s^f)^2$ & 0.0151 & 0.0151 & 0.0126 & 0.0126 & 0.0097 & 0.0097 & 0.0156 & 0.0164 \\ 
%\hspace{0.9cm}
$(c_s^t)^2$ & 0.0159 & 0.0110 & 0.0173 & 0.0104 & 0.0110 & 0.0094 & 0.0146 & 0.0005 \\ 
error in $(c_s^t)^2$ (\%)       & 5.30   & 27.2   & 37.3   & 17.5   & 13.4   &  3.1   & 6.41   & 97.0 \\
\hline
\end{tabular}
\end{center}
\end{table}

\end{document}